\documentclass[aps,prf,superscriptaddress]{revtex4-2}

\usepackage{amsmath,amssymb}
\usepackage{xcolor}
\usepackage{graphicx}
\usepackage{dcolumn}
\usepackage{bm}
\usepackage{siunitx}

\newcommand{\D}{\mathrm{d}}

\newcommand\op[3]{\bm{#1}^{#2}_{#3}}

\newcommand\R[1]{\op{r}{}{#1}}

\newcommand{\rprim}{\bm{r}'}

\newcommand{\rs}{{r^*}}
\newcommand{\giso}{g_0}

\newcommand{\gd}{\dot{\gamma}}
\newcommand{\gtwo}{g}
\newcommand{\gthree}{g_3}

\newcommand{\gSigma}{\bm{\Sigma}}
\newcommand{\gGamma}{\bm{\Gamma}}
\newcommand{\gXi}{\bm{\Xi}}
\newcommand{\gPi}{\bm{\Pi}}
\newcommand{\gUpsilon}{\bm{\Upsilon}}
\newcommand{\gTheta}{\bm{\Theta}}

\newcommand{\gPhi}{\bm{\Phi}}

\newcommand{\pSigma}{\bm{\Sigma'}}

\newcommand{\gF}{\bm{f}}

\newcommand\Msymt[2]{\begin{pmatrix} #1 & #2 \\ #2 & -#1 \end{pmatrix}}

\newcommand{\ror}{\R{}\otimes\R{}}
\newcommand{\rof}{\R{}\otimes\gF(\R{})}
\newcommand\gdot{\bm{\cdot}}

\newcommand{\er}{\bm{e}_r}

\newcommand{\et}{\bm{e}_{\theta}}

\newcommand{\eroer}{\er\otimes\er}

\begin{document}

\title{Microscopically grounded constitutive model for dense suspensions of soft particles below jamming}

\author{Nicolas Cuny}
\affiliation{University of Geneva, Quai Ernest Ansermet 30, 1205 Gen\`eve, Switzerland}
\affiliation{Universit\'e Grenoble Alpes \& CNRS, LIPhy, 38000 Grenoble, France}

\author{Eric Bertin}
\affiliation{Universit\'e Grenoble Alpes \& CNRS, LIPhy, 38000 Grenoble, France}

\author{Romain Mari}
\affiliation{Universit\'e Grenoble Alpes \& CNRS, LIPhy, 38000 Grenoble, France}

\date{\today}

\begin{abstract}
We derive from particle-level dynamics a constitutive model describing the rheology of two-dimensional dense soft suspensions below the jamming transition, in a regime where hydrodynamic interactions between particles are screened. Based on a statistical description of particle dynamics, we obtain through a set of physically plausible approximations a non-linear tensorial evolution equation for the deviatoric part of the stress tensor, involving the strain-rate and vorticity tensors.
This tensorial evolution equation involves singular terms usually not taken into account in phenomenological constitutive models, which most often assume a regular expansion in terms of the stress tensor.
All coefficients appearing in the equation have known expressions in terms of the microscopic parameters of the model. The predictions of this microscopically grounded constitutive model have several qualitative features 
that are specific to the rheology of soft suspensions measured in experiments or simulations. 
The model shows a typical behavior of polymeric visco-elastic materials, such as normal stress differences quadratic in the shear rate $\dot\gamma$, as well as typical behaviors of suspensions of stiff particles, such as a particle pressure linear in $\dot\gamma$ and a zero-shear viscosity diverging at the jamming transition. 
The model also predicts a sharper shear thinning 
than other visco-elastic models at small shear rates, in qualitative agreement with experimental observations.
Furthermore the shear thinning follows a critical scaling close to the jamming transition.
\end{abstract}

\maketitle


\section{Introduction}

Materials made up of soft elastic particles suspended in a fluid form a broad subset of complex fluids.
They include, among others, emulsions, suspensions of microgels, liposomes or vesicles~\cite{bonnecaze_micromechanics_2010}.
In industrial context, they are commonly processed in food or cosmetic industries, but also find applications in e.g. drug delivery \cite{lopez_use_2005} or energy storage \cite{tyagi_development_2011}.
The rheology of these systems is thus of key importance to many industrial processes.
It is also of fundamental interest, being a natural extension of the case of suspensions of hard particles, which can be seen as a limiting case of soft particle suspensions~\cite{nessPhysicsDenseSuspensions2022}.

When the particles are large enough (typically for sizes larger than a micrometer), or under flow when the deformation rate is large enough, these suspensions can be considered athermal, that is, one can neglect the Brownian motion occurring at the particle scale.
The phase diagram of soft athermal systems is well known. 
The key control parameter is the volume fraction of the particle phase, $\phi$.
Below the so-called jamming volume fraction $\phi_\mathrm{J}$, the suspension is a visco-elastic fluid, while above $\phi_\mathrm{J}$, it turns to a yield stress fluid, that is, below a ($\phi$-dependent) yield value of the stress, the suspension is an elastic solid, and above it flows plastically~\cite{bonn_yield_2017}. 
The value of $\phi_\mathrm{J}$ depends on geometrical aspects like particle shape~\cite{nessPhysicsDenseSuspensions2022}, or the nature of constraints to motion created by interactions, such as friction or adhesion~\cite{guy_constraint-based_2018}.
Crucially however, it does not depend on the softness of the particles \emph{per se}, insofar as softness is not affecting friction or adhesion~\cite{gilbert_impact_2022}.

In the fluid phase below jamming, the viscosity $\eta$ of a soft suspension is a function of the volume fraction and the applied shear rate $\dot\gamma$. It is shear-thinning~\cite{pal_rheology_1992,otsubo_rheology_1994, mason_yielding_1996,derkachRheologyEmulsions2009}, and is reasonably well captured by Cross or Carreau-Yasuda-like laws interpolating the viscosity as a function of flow strength between low-stress and high-stress limiting values~\cite{adams_influence_2004,omari_soft_2006,paredes_rheology_2013,dinkgreve_universal_2015}.
This shear-thinning is also captured by numerical simulations~\cite{loewenbergNumericalSimulationConcentrated1996,zinchenkoShearFlowHighly2002,zinchenkoExtensionalShearFlows2015,olsson_critical_2007,olsson_critical_2011,trulsson_athermal_2015,kawasaki_diverging_2015}.
Interestingly, at small $\dot\gamma$ the observed shear thinning is quite steep, with an amplitude $\eta(\dot\gamma)-\eta(0)$ scaling as $\dot\gamma$~\cite{loewenbergNumericalSimulationConcentrated1998,zinchenkoShearFlowHighly2002,zinchenkoExtensionalShearFlows2015,zinchenkoGeneralRheologyHighly2017,panDissipativeParticleDynamics2014} (in this article we consider a definition of $\dot\gamma$ such that $\dot\gamma \geq 0$), or even $\dot\gamma^y$ with $y<1$~\cite{paredes_rheology_2013,dinkgreve_universal_2015,vagbergUniversalityJammingCriticality2014,kawasaki_diverging_2015}, when 
usual visco-elastic models of the upper-convected Maxwell family (e.g., Oldroyd model, Giesekus model, Phan-Thien-Tanner model~\cite{macoskoRheologyPrinciplesMeasurements1994}) all predict $\eta(\dot\gamma)-\eta(0) \sim \dot\gamma^2$.

In a simple picture, shear thinning is a consequence of the fact that when the applied shear rate increases, stresses lead to increasing particle deformation, so that particles can better accommodate the applied flow.
This can be interpreted in terms of an effective volume fraction $\phi_\mathrm{eff} \lesssim \phi$ decreasing with increasing applied stress, 
such that the viscosity is well approximated as $\eta_\mathrm{Hard}(\phi_\mathrm{eff})$, where $\eta_\mathrm{Hard}(\phi)$ is the viscosity for a suspension of hard spheres at volume fraction $\phi$~\cite{trulsson_athermal_2015,rostiRheologySuspensionsViscoelastic2018,aouaneStructureRheologySuspensions2021}.
Beyond this simple qualitative picture, however, there is to the best of our knowledge currently no theory aiming at capturing the essential features of shear thinning for athermal suspensions of soft particles.

Soft suspensions also show non-trivial normal stresses~\cite{loewenbergNumericalSimulationConcentrated1996,loewenbergNumericalSimulationConcentrated1998,zinchenkoShearFlowHighly2002,zinchenkoExtensionalShearFlows2015,zinchenkoGeneralRheologyHighly2017,malekiViscousResuspensionDroplets2022}. In the small shear rate limit, normal stress differences scale as $\dot\gamma^2$~\cite{loewenbergNumericalSimulationConcentrated1996,zinchenkoShearFlowHighly2002} in a typical visco-elastic fluid fashion (although sometimes scaling in $\dot\gamma$ is observed~\cite{clausenRheologyMicrostructureConcentrated2011,panDissipativeParticleDynamics2014}).
They are found to be of opposite signs, $N_1>0$ and $N_2<0$, with $N_1 > |N_2|$.
However, taken individually each normal stress scales as $\dot\gamma$~\cite{zinchenkoGeneralRheologyHighly2017,malekiViscousResuspensionDroplets2022}, just like for a suspension of hard particles~\cite{dennRheologyNonBrownianSuspensions2014}.

Existing constitutive models of soft dense suspensions are struggling to capture the observed phenomenology.
Constitutive models have been developed in the dilute regime, based on single particle dynamics~\cite{schowalterRheologicalBehaviorDilute1968,frankelConstitutiveEquationDilute1970,barthes-bieselRheologySuspensionsIts1973} or semi-dilute regime, based on particle-pair dynamics~\cite{choiRheologicalPropertiesNondilute1975,zinchenkoEffectHydrodynamicInteractions1984}, but these are limited to low volume fractions as they are predicting the stress to order $\phi$ or $\phi^2$, respectively.
Doi-Ohta theory for mixtures of immiscible fluids~\cite{doiDynamicsRheologyComplex1991} predicts no shear thinning and normal stress differences linear in $\dot\gamma$ at small $\dot\gamma$~\cite{guentherEvaluationDoiOhta1998}.
Only the generalized Oldroyd model with stress and rate dependent coefficients by Martin et al.~\cite{martinGeneralizedOldroydModel2014} is accurately capturing the observed rheology, but this requires to promote the coefficients to arbitrary functions of an invariant (the double contraction of the stress and strain-rate tensors) which are then tabulated from observations.

Here, we introduce a microscopic theory of shear thinning in suspensions of soft particles. 
It is based on an approach we recently introduced for deriving constitutive equations for dense soft particle systems~\cite{CunyPRL21,CunyJSTAT22}. 
In these works, we obtained a constitutive equation for a minimal model of soft jammed suspensions~\cite{durian_foam_1995}.
A key step to this approach is the use of a closure of the microstructure, via the pair correlation function, as a function of the stress tensor of the suspension. 
In~\cite{CunyPRL21,CunyJSTAT22}, this closure was taylored to the jammed phase.
Here, we revisit this closure in the case of the flowing regime, below jamming. 
We then obtain a non-linear visco-elastic evolution equation for the deviatoric part of the stress tensor, which coefficients are explicitly related to suspension properties.
This constitutive model exhibits shear thinning, and  predicts that the particle pressure and the shear thinning amplitude are both linear in the absolute value of the deformation rate at leading order, while normal stress differences are quadratic in the deformation rate.

\section{Soft suspension model}
\label{sec:model}

We adopt as our microscopic model of a two-dimensional soft suspension the Durian model~\cite{durian_foam_1995,durianBubblescaleModelFoam1997} that consists of $N$ soft disks of radius $a$, with overdamped and athermal dynamics.
Particles interact only via radial repulsion forces, and gravity is not taken into account.
Particles experience a viscous drag resulting from the fluid they are implicitly immersed in. The back action of the particles on the fluid is neglected.
Under these simplifying assumptions, the fluid can be characterized by an affine velocity field $\bm{u}(\bm{r})$.
The fluid velocity gradient is assumed uniform, $\bm{u}(\bm{r}) = \nabla \bm{u} \gdot \bm{r}$
(we use the convention $(\nabla\bm{u})_{ij}=\partial u_i/\partial r_j$ to define the gradient of the vector field $\bm{u}$), 
and we define the strain-rate tensor $\bm{E} = (\nabla \bm{u} + \nabla \bm{u}^\mathrm{T})/2$, the vorticity tensor $\bm{\Omega} = (\nabla \bm{u} - \nabla \bm{u}^\mathrm{T})/2$, and the shear rate $\dot\gamma = \sqrt{2\bm{E}:\bm{E}}$ (with the double contraction of two tensors $\bm{A}$ and $\bm{B}$ defined as $\bm{A}:\bm{B} = \sum_{ij} A_{ij} B_{ij}$).
The particle density $\rho=N/V$, where $V$ is the volume (actually an area in two dimensions) is also assumed to be uniform.
The position of particle $\mu$ is denoted as $\op{r}{}{\mu}$, and its velocity is denoted as $\dot{\bm{r}}_\mu$.
To lighten notations, it is convenient to write $\bm{u}_{\mu}=\bm{u}(\bm{r}_{\mu})$ the velocity field of the fluid at the position $\bm{r}_{\mu}$ occupied by particle $\mu$.
The viscous drag exerted by the fluid on particle $\mu$ is then equal to $-\lambda_{\rm f} (\dot{\bm{r}}_\mu-\bm{u}_{\mu})$, where $\lambda_{\rm f}$ is the viscous friction coefficient.
The pairwise repulsive contact force exerted by particle $\nu$ on particle $\mu$ is given by
$\bm{f}(\op{r}{}{\mu\nu}) = f(r_{\mu\nu})\op{r}{}{\mu\nu}/r_{\mu\nu}$, with $\op{r}{}{\mu\nu}=\op{r}{}{\nu}-\op{r}{}{\mu}$, $r_{\mu\nu} = |\op{r}{}{\mu\nu}|$. The case of a repulsive force corresponds to $f(r) \le 0$.
We keep the contact force generic at this stage, only assuming that $f(r_{\mu\nu}) = 0$ for $r_{\mu\nu} > 2a$ by definition of the contact.
Yet later on to perform explicit calculations we will assume repulsive harmonic disks, that is 
$f(r)=f_0 (r/a-2)$ for $r<2a$ and $f(r)=0$ for $r>2a$.

Calling more generically $f_0$ a typical contact force, we work with dimensionless variables, using an elastic unit system with a unit force $f_0$, a unit time $\tau_0=\lambda_{\rm f} a/(2f_0)$ (corresponding to the elastic relaxation time), and a unit length $a$.
In terms of dimensionless variables (indicated here with a hat), the equation of motion of particle $\mu$ reads
\begin{equation}
    \label{dynamics}
    -2 (\hat{\dot{\bm{r}}}_\mu - \hat{\bm{u}}_{\mu})+ \sum_{\nu (\ne\mu)} \hat{\bm{f}}(\hat{\bm{r}}_{\mu\nu}) = \bm{0}.
\end{equation}
In the elastic unit system we picked, the dimensionless shear rate $\hat{\dot\gamma}$ is nothing but the Weissenberg number (or the capillary number if the elastic force is of interfacial origin).
In consequence, the limit $\hat{\dot\gamma}\to 0$ is the hard-sphere limit of our model.
In the following we drop the hat on dimensionless variables to lighten notations.

\section{Stress tensor dynamics}

\subsection{Exact evolution equation for the stress tensor}

We wish to derive an evolution equation for $\gSigma$, the elastic contribution to the stress tensor of the suspension.
The elastic stress tensor $\gSigma$ is defined in terms of the pair correlation function $g(\R{})$ (i.e., the probability to find a particle at a position $\R{}$ with respect to a given particle) characterizing the microstructure of the suspension, by the Virial formula \cite{nicot_definition_2013}
\begin{equation}
    \label{virial}
    \gSigma=\frac{\rho^2}{2}\int\left(\rof\right) g(\R{})\D\R{},
\end{equation}
where $\D\R{}$ denotes the two-dimensional integration element.
In two dimensions, the trace of the stress tensor $\gSigma$ is equal to $-2p$, where $p$ is the pressure, so that from the expression \eqref{virial} of the stress tensor, the pressure $p$ is given by
\begin{equation}
	\label{def:pressure}
        p = -\frac{\rho^2}{4}\int r f(r) \gtwo(\R{})\D\R{}\,.
\end{equation}
An exact, but not closed, evolution equation for $\gSigma$ has been derived in \cite{CunyPRL21,CunyJSTAT22}, starting from the evolution equation of the pair correlation function $g(\R{})$.
In the following, we focus on the evolution equation for the deviatoric (i.e., traceless) part $\gSigma'=\gSigma-\frac{1}{2}{\rm Tr}(\gSigma)\,\bm{1}$ of the stress tensor, which reads
\begin{equation}
\frac{\mathrm{D} \gSigma'}{\mathrm{D} t} = \bm{\Theta}' - \bm{\Phi}' - \bm{\Xi}' - \bm{\Pi}' - \bm{\Gamma}' - \bm{\Upsilon}',
        \label{eq:sigmaprim}
\end{equation}
where we have used the upper-convected Maxwell derivative defined as
\begin{equation} \label{eq:def:UCM:derivative}
\frac{\mathrm{D} \gSigma'}{\mathrm{D} t} \equiv \dot{\gSigma}' - \bm{\Omega} \gdot\, \gSigma' + \gSigma' \gdot\, \bm{\Omega} + 2p \bm{E}\,,
\end{equation}
in its form suited for the traceless tensor $\gSigma'$ (the form would slightly differ for the time derivative of $\gSigma$, which is often used in the literature). Note that this specific form of the material derivative of $\gSigma'$ is imposed by frame indifference as long as one considers systems with overdamped dynamics.
The tensors $\gTheta',\dots,\gPhi'$ appearing in the rhs of
Eq.~\eqref{eq:sigmaprim} are the deviatoric parts of the following tensors:
\begin{align}
        \label{def_Theta}
        &\bm{\Theta} = \frac{\rho^2}{2} \int \left(\bm{E}:\eroer\right) (\ror)\gdot\nabla\gF\, g(\R{})\D\R{}, \\
        \label{def_Phi}
        &\bm{\Phi} = \frac{\rho^2}{2} \int \left(\bm{E}:\eroer\right) (\rof)\, g(\R{})\D\R{}, \\
        \label{def_Xi}
        &\bm{\Xi}= \frac{\rho^2}{2} \int \left(\gF(\R{})\otimes\gF(\R{})\right)g(\R{})\D\R{}, \\
        \label{def_Pi}
        &\bm{\Pi}=\frac{\rho^2}{2} \int\left(\R{}\otimes\gF(\R{})\right)\gdot\,\nabla\gF^\mathrm{T} g(\R{})\D\R{},\\
        \label{def_Gamma}
        &\bm{\Gamma}=\frac{\rho^3}{2}\iint \left(\gF(\rprim)\otimes \gF(\R{})\right) \gthree(\R{},\rprim)\D\R{}\D\rprim, \\
        \label{def_Upsilon}
        &\bm{\Upsilon}=\frac{\rho^3}{2} \iint \left(\R{} \otimes \gF(\rprim)\right)\gdot \left(\nabla\gF(\R{})\right)^\mathrm{T}\gthree(\R{},\rprim)\D\R{}\D\rprim,
\end{align}
where $\gthree(\R{},\rprim)$ is the three-body correlation function (i.e., the probability to find two particles respectively at positions
$\R{}$ and $\rprim$ with respect to a given particle situated at the origin).
Up to this point, the evolution equation \eqref{eq:sigmaprim} for $\gSigma'$ is exact and is the same as the one considered above the jamming density in \cite{CunyPRL21,CunyJSTAT22}.

However, Eq.~\eqref{eq:sigmaprim} is not a closed evolution equation for the deviatoric stress tensor $\gSigma'$, as it involves the pair and three-body correlation functions. To close this equation, we use the same strategy as above the jamming density in \cite{CunyPRL21,CunyJSTAT22}, which consists in three successive approximation steps.
First, we approximate the three-body correlation function $\gthree$ in terms of the pair correlation function $\gtwo$ using the simple Kirkwood closure \cite{kirkwood_statistical_1935}:
\begin{equation} \label{eq:Kirkwood}
\gthree(\R{},\rprim) = \gtwo(\R{}) \, \gtwo(\rprim) \, \gtwo(\R{}-\rprim)\,.
\end{equation}
Second, we introduce using plausible physical arguments an approximate expression of the anisotropic pair correlation function in terms of the deviatoric stress tensor $\gSigma'$ and of an isotropic pair correlation function, focusing on the weakly anisotropic limit.
Finally, we devise a minimal parametrization of the isotropic pair correlation function, leaving no free parameters in the description.
These two last approximation steps are described in the next section.
Quite importantly, the difference between situations above and below the jamming density mainly comes from the parametrization of the pair correlation function, as discussed below.

\subsection{Parametrization of the pair correlation function}

\begin{figure}
\includegraphics[width=0.4\columnwidth]{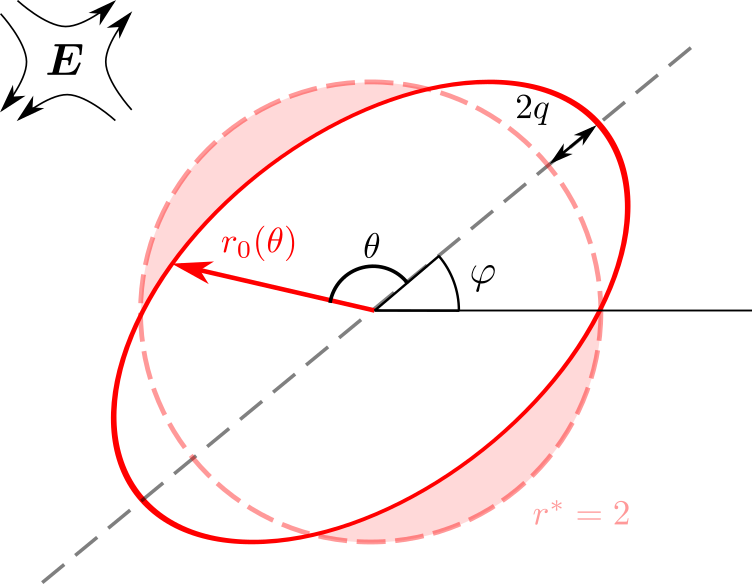}
\caption{Schematic representation of the deformation of the first neighbor shell below the jamming density.}
\label{fig:deformed_ring_under_jamming}
\end{figure}

We assume that for densities close enough but below the jamming density, neighboring particles are just in contact. Under weak deformation, we further assume that the ring of first neighbors deforms into a second order harmonic, in a way analogous to the assumptions made in \cite{CunyPRL21,CunyJSTAT22}.
To parametrize this deformation, we introduce two scalar parameters: $q$ the amplitude of the deformation and $\varphi$ the orientation of the extensional axis of the microstructure.
Using polar coordinates $(r,\theta)$ and choosing $\varphi$ as the origin of the angles (i.e., $\theta=0$, see Fig.~\ref{fig:deformed_ring_under_jamming}), we parametrize the ring of first neighbors by
\begin{equation}
        \label{r0_under_jamming}
        r_0(\theta)=2\left(1+q\cos2\theta\right).
\end{equation}
As in the case of suspensions above the jamming density \cite{CunyPRL21,CunyJSTAT22}, we assume that the pair correlation function is deformed homothetically to the ring of first neighbors with respect to $\giso$, its isotropic pair correlation in the absence of deformation:
\begin{equation}
        \label{param_g}
        \gtwo(\R{})=\giso\left(\frac{r}{1+q\cos2\theta}\right),
\end{equation}
with $r=|\R{}|$. 
We then need to parametrize the isotropic pair correlation function $\giso(r)$.
It is possible to use the same type of schematic description for $\giso(r)$ as in \cite{CunyPRL21,CunyJSTAT22},
\begin{equation}
        \label{giso_under_jamming}
        \giso(r)=A\, \delta(r-2)+H(r-2),
\end{equation}
where the delta peak $A\,\delta(r-2)$ schematically represents the first shell of neighbors situated at $r=2$, and the Heaviside function $H(r-2)$ describes the sea of neighbors situated beyond the first shell, neglecting secondary peaks (the Heaviside function is defined as $H(x)=1$ for $x>0$ and $H(x)=0$ for $x\le 0$).
As a minimal hypothesis, we assume that close to the jamming transition, the first shell of neighbors contains approximately $z=6$ particles, which leads to an amplitude $A=z/(4\pi \rho)$ of the delta peak, using the fact that $z$ is the integral of $\rho \giso(r)$ over a thin circular shell around $r=2$.
A key difference with the situation above jamming is that in the latter case, the first shell of neighbors is situated at $r=\rs<2$, while below (but close to) jamming the position of the first shell of neighbors remains fixed at $r=2$.

The parametrization \eqref{param_g}, \eqref{giso_under_jamming} of the pair correlation function has important consequences on the evaluation of the integrals defining the tensors $\gTheta,\dots,\gUpsilon$, see Eqs.~\eqref{def_Theta} to \eqref{def_Upsilon}.
Integrals involving $\gtwo(\R{})$ over the domain $\mathcal{C}=\left\{\R{},\, |\R{}|<2\right\}$ boil down to integrals over a reduced angular domain since any $\theta$ direction such that $r_0(\theta)>2$ does not contribute to the integral. Only the quadrants $\left[-3\pi/4,-\pi/4\right]$ and $\left[\pi/4,3\pi/4\right]$ have a non-zero contribution.
This is a key difference with polymeric models, for which all quadrants contribute to the stress.

\subsection{Evaluation of the stress tensor}
\label{sec:stress-tensor}

\subsubsection{Deviatoric part of the stress tensor}

Using the Virial definition \eqref{virial} of the stress tensor $\gSigma$ as well as the parametrization~\eqref{param_g} of $\gtwo(\R{})$, one may write the deviatoric part $\gSigma'$ of the stress tensor as
\begin{equation}
        \gSigma' = \frac{\rho^2}{4}\int_{-\pi}^\pi \D\theta  \int_0^2 \D r\, r^2f(r) \,\giso\!\left(\frac{r}{1+q\cos2\theta}\right) \, \Msymt{\cos2\theta}{\sin2\theta},
        \label{def_sigmap}
\end{equation}
using for calculations purposes the basis defined by the elongated and compressed axes of the microstructure.
The antisymmetry in $\theta$ of the off-diagonal coefficients of the integrand implies that $\gSigma'$ is diagonal in this basis.
Using Eqs.~(\ref{param_g}) and (\ref{giso_under_jamming}) as well as the symmetry of the integrand,
one can reduce the angular integration interval to $\left[\pi/4,3\pi/4\right]$.
Making the change of variable $\tilde{r}=r/(1+q\cos2\theta)$, we can then calculate $\gSigma'$, whose exact form is a quartic polynomial in $q$ times a diagonal traceless tensor.
In what follows, we aim at deriving a minimal model and expand $\gSigma'$ up to linear order in $q$, leading to
\begin{equation}
        \label{calcul_sigmaq}
        \gSigma' = \pi  A \rho^2 q  \Msymt{1}{0}.
\end{equation}
If this expression seems to suggest that the principal axes of the $\gSigma'$ tensor do not vary in time, it must be kept in mind that Eq.~(\ref{calcul_sigmaq}) is the expression of $\gSigma'$ in the basis such that the elongated axis of the microstructure is the origin of the angles. In practice, this basis can itself vary with time. The expression of $\gSigma'$ tells us that its principal axes are aligned with the elongated and compressed axes of the microstructure. 
A similar relation between stress tensor and microstructure was found above the jamming density \cite{CunyPRL21,CunySM22}.

For later convenience, we further note that from Eq.~\eqref{calcul_sigmaq}
the norm $|\gSigma'| = \sqrt{\gSigma':\gSigma'/2}$ of the tensor $\gSigma'$ is given to first order in $q$ by
\begin{equation}
        \label{calcul_sigmaq_norm}
        |\gSigma'| = \pi A \rho^2 q \,.
\end{equation}
This relation shows that the stress amplitude is directly related to the anisotropy of the microstructure.

\subsubsection{Evaluation of the pressure}

Although we focus on the derivation of an evolution equation for the deviatoric part $\gSigma'$ of the stress tensor, we also need to evaluate the pressure $p$ because it appears in the r.h.s.~of the evolution equation \eqref{eq:sigmaprim}, as a prefactor of the tensor $\bm{E}$.
Using the weakly anisotropic parametrization \eqref{param_g} of the pair correlation function in the integral \eqref{def:pressure} defining the pressure $p$, we can again reduce the angular integration interval to $\left[\pi/4,3\pi/4\right]$, yielding
\begin{equation}
        \label{calcul_p}
        p = -\frac{\rho^2}{2}\int_{\pi/4}^{3\pi/4}\D\theta\int_0^2 \D r\, r^2f(r)\,\giso\!\left(\frac{r}{1+q\cos2\theta}\right).
\end{equation}
Using the change of variable $\tilde{r}=r/(1+q\cos2\theta)$, and expanding the resulting expression to first order in $q$, we get
\begin{equation}
	\label{eq:pressure:q}
	p = 4A \rho^2 q \,.
\end{equation}
%
Using Eq.~\eqref{calcul_sigmaq_norm}, one can then reexpress the pressure $p$ as a function of the norm
$|\gSigma'|$ of the deviatoric stress tensor, leading to
\begin{equation}
	\label{eq:pressure:sigma}
	p = \frac{4}{\pi} \, |\gSigma'| \,.
\end{equation}
Note that the amplitude $A$ also disappears from the relation between $p$ and $ |\gSigma'|$, which includes only fixed numerical prefactors.
As explained below, Eq.~\eqref{eq:pressure:sigma} plays an important role in the rheological behavior of the model, as in implies that the pressure is proportional to the shear rate in the low shear-rate limit.

\subsection{Closed evolution equation on the deviatoric stress tensor}
\label{sec:closed}

We are now in a position to evaluate the tensorial integral terms $\bm{\Theta}$, $\bm{\Phi}$, $\bm{\Xi}$, $\bm{\Pi}$, $\bm{\Gamma}$ and $\bm{\Upsilon}$ defined in Eqs.~\eqref{def_Theta} to \eqref{def_Upsilon}, thanks to the weakly anisotropic parametrization \eqref{param_g} of $\gtwo(\R{})$ and the schematic parametrization \eqref{giso_under_jamming} of $\giso(r)$.
Calculations are made for the specific force $f(r)=r-2$, corresponding to repulsive harmonic disks.
Keeping calculations at lowest order, all integrals are evaluated by performing an expansion to order $q$ in the weakly anisotropic limit ($q\ll 1$).
Details of the derivations can be found in Appendix~\ref{app:integrals}.
The calculation is actually easy for the tensor $\bm{\Pi}$, for which we find the exact result
$\bm{\Pi}'=\gSigma'$.
For the other tensorial integrals, we obtain after expansion to order $q$:
\begin{align}
        \label{def_Theta_exp}
        	\bm{\Theta}' &= \pi A \rho^2 \bm{E} \!- \frac{4(2A-1)}{3\pi A} \left( 2|\gSigma'| \, \bm{E}
	\!+ \frac{\gSigma':\bm{E}}{|\gSigma'|} \, \gSigma' \right)\!,\\
        \label{def_Phi_exp}
	\bm{\Phi}' &= -\frac{2}{3\pi} \left( 2|\gSigma'| \, \bm{E}
	+ \frac{\gSigma':\bm{E}}{|\gSigma'|} \, \gSigma' \right)\!,\\
        \label{def_Upsilon_exp}
	\bm{\Upsilon}' &= -\frac{B \rho}{9\pi}\, \gSigma',
\end{align}
with
\begin{equation} \label{eq:coef:B}
B = 47A-9A^2+A(A+3)\pi \sqrt{3}
\end{equation}
[we recall that $A=3/(2\pi\rho)$].
The tensorial integrals $\bm{\Xi}'$ and $\bm{\Gamma}'$ have a leading contribution at order $q^2$ only, and can thus be neglected at order $q$.
It follows that a closed evolution equation can be written for the deviatoric part $\gSigma'$ of the stress tensor,
%
\begin{equation}
        \label{eq:sigmaprim2}
        \frac{\mathrm{D} \gSigma'}{\mathrm{D} t} = \left( \kappa - \lambda |\gSigma'| \right) \bm{E}
        - \left( \beta + \xi \frac{\gSigma':\bm{E}}{|\gSigma'|} \right) \gSigma'
\end{equation}        
%
Note that the pressure $p$ that appears in the definition (\ref{eq:def:UCM:derivative}) of the upper-convected Maxwell derivative may be expressed as a function of $|\gSigma'|$ thanks to Eq.~\eqref{eq:pressure:sigma}.
The coefficients appearing in Eq.~\eqref{eq:sigmaprim2} are explicitly given in terms of microscopic parameters as
\begin{align}
	\kappa &= \pi A \rho^2 \,,\\
	\lambda &= \frac{4(3A-2)}{3\pi A} \,,\\
 \label{eq:def:beta}
	\beta &= 1-\frac{B \rho}{9\pi} \equiv \frac{\beta_0}{\phi} \, (\phi_\mathrm{J}-\phi) \,,\\
	\xi &= \frac{2(3A-2)}{3\pi A}  \,,
\end{align}
with $\phi=\pi\rho$ the packing fraction, and where we have defined 
\begin{equation}
    \phi_\mathrm{J} = \frac{3(9-\pi\sqrt{3})}{2(47+3\pi\sqrt{3}-6\pi^2)}\,, \qquad
    \beta_0 = \frac{47+3\pi\sqrt{3}}{6\pi^2}-1\,,
\end{equation}
where $\phi_\mathrm{J}$ is to be interpreted as the jamming packing fraction.
Indeed, in the absence of applied strain, $\bm{E}=0$, the stress tensor should relax to zero, as we are dealing with a dense suspension below the jamming density. 
One should thus have $\beta>0$, as the coefficient $\beta$ governs the linear stability of the state $\gSigma=0$.
The coefficient $\beta$ is a decreasing function of the density $\rho$, or equivalently of the packing fraction $\phi = \pi\rho$, and it vanishes for a packing fraction $\phi_{\rm J} \approx 1.30$, which we thus identify as the jamming volume fraction. 
Its value is slightly larger than the one obtained in the case above jamming in \cite{CunyPRL21,CunyJSTAT22}, and approximately $50\%$ larger than the correct packing jamming fraction in two dimensions, due to the approximations made.
On the other hand, we have neglected in the derivation the effect of the hydrodynamic interactions between particles, which is justified only in the dense regime.
The regime of validity of our derivation is thus limited to packing fractions $\phi$ close to, but below, $\phi_{\rm J}$.
It is also interesting to note that the $\rho$-dependence of the coefficient $\beta$ as given in Eq.~\eqref{eq:def:beta} comes from three-body correlations (through the tensorial integral $\bm{\Upsilon}'$), while the other coefficients $\kappa$, $\lambda$ and $\xi$ only result from pair correlations, through the interplay of pairwise repulsion with the applied flow.

Our constitutive equation \eqref{eq:sigmaprim2} is akin to the upper-convected Maxwell (UCM) equation~\cite{oldroydFormulationRheologicalEquations1950}, where the upper convective derivative of the stress tensor (or, in our case, the traceless part of this derivative) is expressed as a function of the stress tensor itself and of the strain rate tensor. 
Models of the UCM type, used to describe visco-elastic fluids, are many~\cite{macoskoRheologyPrinciplesMeasurements1994,larsonConstitutiveEquationsPolymer1988}.
They typically involve the same tensorial terms as the r.h.s.~of Eq.~\eqref{eq:sigmaprim2}, that is, terms of the form $f_1(\gSigma', \bm{E}) \gSigma'$ and $f_2(\gSigma', \bm{E}) \bm{E}$ (sometimes alongside other terms). 
The functions $f_1(\gSigma', \bm{E})$ and $f_2(\gSigma', \bm{E})$ are scalar functions of the simultaneous invariants under orthogonal transformations of the two tensors $\gSigma'$ and $\bm{E}$, following the Hand framework~\cite{hand_theory_1962} for frame-indifferent dynamics of a symmetric second-rank tensor.
In our case, as $\gSigma'$ and $\bm{E}$ are two-dimensional traceless tensors, these invariants are $\gSigma':\gSigma'$, $\bm{E}:\bm{E}$ and $\gSigma':\bm{E}$.
Now, many UCM-like models assume $f_1$ and $f_2$ to be simple analytical functions. 
Our derivation, based on an expansion in the stress anisotropy amplitude, also reveals simple functional forms for $f_1$ and $f_2$, albeit rather ``singular'' ones, as they involve the tensorial norm $|\gSigma'| = \sqrt{\gSigma':\gSigma'/2}$, which to our knowledge is unique to our model.
The other possible invariants appear in some UCM-like models, e.g.~$\gSigma':\bm{E}$ appears in the Larson model~\cite{larsonConstitutiveEquationPolymer1984}, and $\bm{E}:\bm{E}$ appears in the White and Metzner model~\cite{whiteDevelopmentConstitutiveEquations1963}.
We discuss below the consequences of the presence of these singular terms on the rheology of dense soft suspensions.

Intriguingly, these ``singular'' terms have closely related analogues in 
the semi-phenomenological Doi-Ohta theory for mixtures of immiscible fluids~\cite{doiDynamicsRheologyComplex1991} (which include, but are not limited to emulsions).
The central outcome of Doi-Ohta theory is a coupled time evolution for the so-called interface tensor $\bm{q}$, which is the traceless second moment tensor of the distribution of unit normals on a droplet interface deformed by the flow, and the interface area $Q$.
The singular terms in the stress evolution Eq.~\eqref{eq:sigmaprim2} all have equivalents in Doi-Ohta theory provided we perform the substitutions $\pSigma \leftrightarrow \bm{q}$ and $|\pSigma| \leftrightarrow Q$.
An additional term proportional to $Q\bm{q}$ present in the Doi-Ohta theory would also appear in our constitutive model as a term in $|\gSigma'| \gSigma'$ by performing the weakly anisotropic expansion up to order $q^2$. 
There are however two crucial differences.
First, in our approach these terms are relaxation terms coming from interactions, unlike in Doi-Ohta theory where they result from the closure of the advection of $\bm{q}$ (although phenomenological extensions of the theory include relaxation terms induced by surface tension with a similar form~\cite{leeRheologyDynamicsImmiscible1994,bousminaRheologyPolymerBlends2001a}).
Second, in Doi-Ohta theory, $Q$ has its own dynamics, and is not proportional to $|\bm{q}|$. Instead, in steady state, one has $Q^2 \propto - \bm{E}:\bm{q}$~\cite{doiDynamicsRheologyComplex1991}.

\section{Rheological properties}


\subsection{Steady-state rheology, shear thinning}

\begin{figure}
\includegraphics[width=\columnwidth]{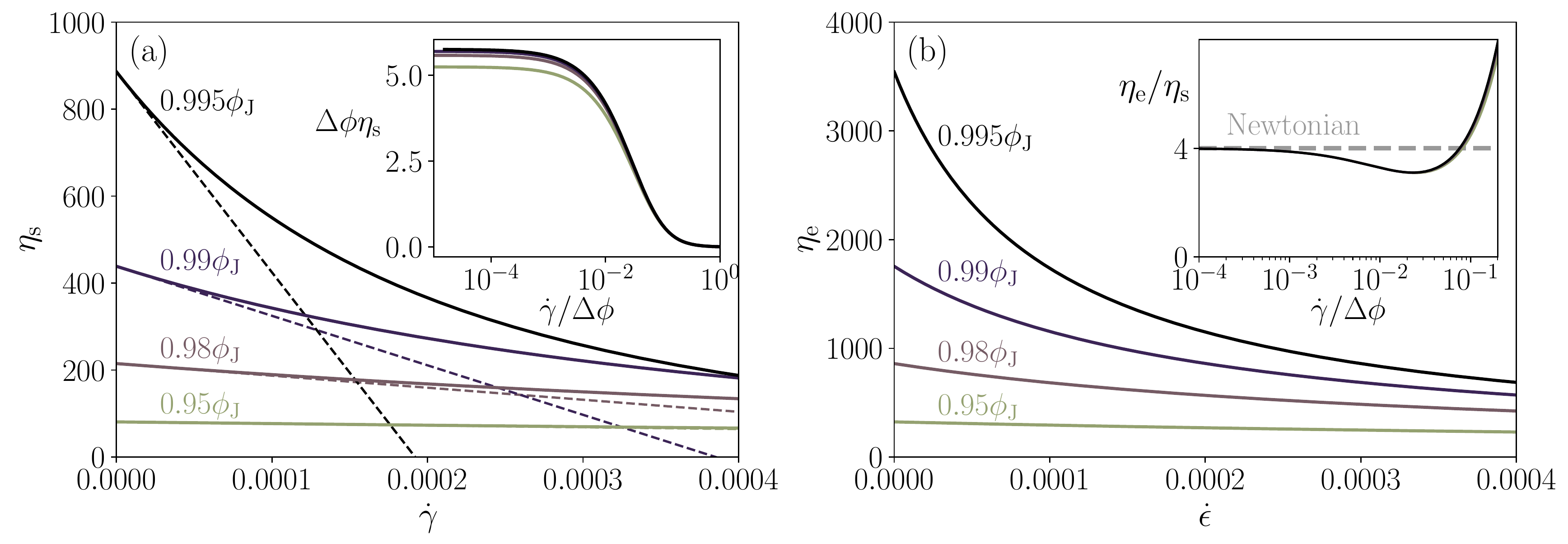}
\caption{Flow curves in simple shear (a) and planar elongational flow (b). (a) Dimensionless viscosity $\eta_\mathrm{s}$ as a function of dimensionless shear rate $\dot\gamma$ in logarithmic scale for several volume fractions below the jamming volume fraction $\phi_\mathrm{J}$. 
In inset, scaled flow curves $\Delta\phi \eta_\mathrm{s}$ as a function of scaled shear rate $\dot\gamma/ \Delta\phi$, evidencing the critical scaling of the viscosity when approaching jamming.  
(b) Extensional viscosity $\eta_\mathrm{e}$ as a function of extension rate $\dot\epsilon$. Inset: Trouton ratio $\eta_\mathrm{e}(\dot\epsilon)/\eta_\mathrm{s}(\dot\gamma)$ as a function of the scaled deformation rate, for $\dot\epsilon=\dot\gamma$. Trouton ratio takes its Newtonian value in the small rate limit, and for larger rates is predicted alternatively sub-Newtonian and super-Newtonian.}
\label{fig:flow_curves}
\end{figure}

We first investigate the steady-state rheology predicted by our model.
We get the steady-state solution of Eq.~\eqref{eq:sigmaprim2} numerically and show the obtained flow curves in Fig.~\ref{fig:flow_curves}, in two cases,  simple shear $\nabla \bm{u} = \dot\gamma \bm{e}_1 \otimes \bm{e}_2$ for which the shear viscosity is defined as $\eta_\mathrm{s} = \Sigma_{12}/\dot\gamma$, and planar extensional flow $\nabla \bm{u} = \dot\epsilon (\bm{e}_1 \otimes \bm{e}_1 - \bm{e}_2 \otimes \bm{e}_2)$ for which we define the extensional viscosity $\eta_\mathrm{e} = (\Sigma_{11} - \Sigma_{22})/\dot\epsilon$.
In both cases, we observe a shear-thinning behavior, which we can characterize analytically in the limit of small deformation rates.
The Trouton ratio $\mathrm{Tr} = \eta_\mathrm{e}/\eta_\mathrm{s}$ of extensional and shear viscosities evaluated at the same deformation rates $\dot\epsilon = \dot\gamma$, shown in the inset of Fig.~\ref{fig:flow_curves}b, is predicted to be taking the Newtonian value $\mathrm{Tr} = 4$ in the limit of small deformation rates. 
At intermediate rates, $\mathrm{Tr} < 4$, in contrast to the high shear rates for which our model predicts super-Newtonian values  $\mathrm{Tr} > 4$.

To get analytical results, we evaluate perturbatively the viscosity at low strain rate. To keep the calculation generic, without explicit reference to the flow geometry, we use the strain rate $\dot\gamma=2|\bm{E}|$. 
For an extensional flow, we thus have $\dot\gamma=2 \dot\epsilon$.
Expanding the steady-state stress in powers of the shear rate $\dot\gamma$ as $\gSigma'_\mathrm{st} = \bm{\Sigma}_1 \dot\gamma + \bm{\Sigma}_2 \dot\gamma^2 + o(\dot\gamma^2)$, we get from  Eq.~\eqref{eq:sigmaprim2}, defining $\hat{\bm{E}} = \bm{E}/\dot\gamma$ and $\hat{\bm{\Omega}} = \bm{\Omega}/\dot\gamma$,
\begin{align}
    \bm{\Sigma}_1 & = \frac{\kappa}{\beta} \hat{\bm{E}}\, ,\label{eq:first_order_stress} \\
    \bm{\Sigma}_2 & = -\frac{\kappa}{2\beta^2} \left(\tilde{\lambda}  + 2\xi \right) \hat{\bm{E}} + \frac{1}{\beta} \left[\hat{\bm{\Omega}} \gdot \hat{\bm{E}} - \hat{\bm{E}}\gdot \hat{\bm{\Omega}} \right]\, ,\label{eq:second_order_stress}
\end{align}
with $\tilde{\lambda}=\lambda+\frac{8}{\pi}$ (the coefficient $\tilde{\lambda}$ gathers the contributions of terms proportional to $|\gSigma'|\bm{E}$ in Eq.~\eqref{eq:sigmaprim2}, coming from the $\lambda$ term and from the pressure term in the upper-convected Maxwell derivative).
The viscosity $\eta = \bm{\Sigma}'_\mathrm{st}:\hat{\bm{E}}/\dot\gamma$ is then
\begin{equation}
    \eta = \frac{\kappa}{2\beta} \left[ 1 - \frac{\tilde{\lambda} + 2\xi}{2 \beta} \dot\gamma \right] + o(\dot\gamma)\, ,
\label{eq:shear_thinning_expansion}\end{equation}
where we used $(\hat{\bm{\Omega}}\cdot \hat{\bm{E}} - \hat{\bm{E}}\cdot \hat{\bm{\Omega}}):\hat{\bm{E}} = 0$ and $|\hat{\bm{E}}| = 1/2$.

\begin{figure}
\includegraphics[width=0.6\columnwidth]{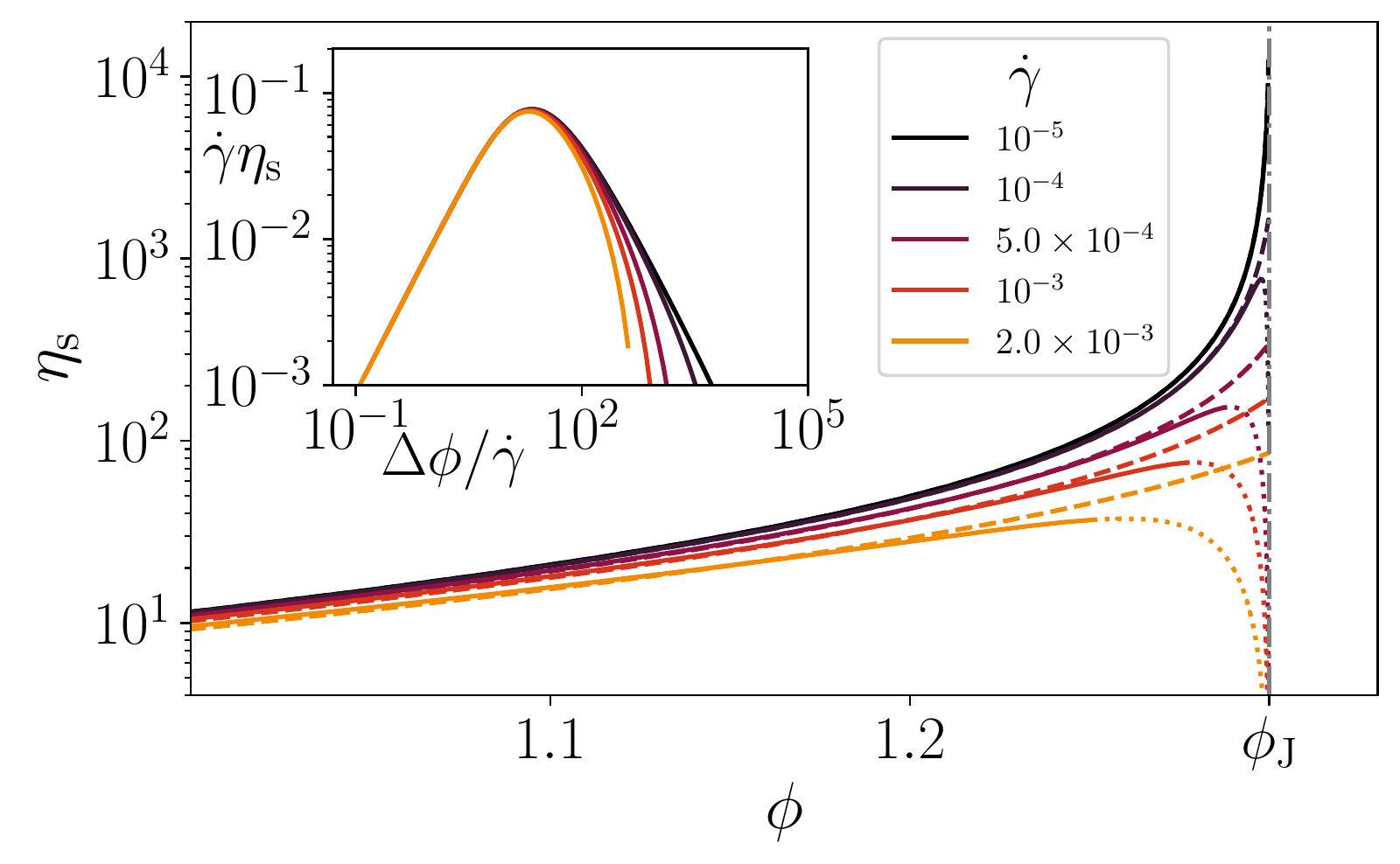}
\caption{Viscosity $\eta_\mathrm{s}$ in simple shear as a function of the volume fraction $\phi$ for several values of $\dot\gamma$ (solid lines) increasing from dark to light colors). The viscosity decreases close for large $\dot\gamma/\Delta\phi$ values beyond the domain of validity of the model (dotted lines).
In dashed lines, we show the approximation with the effective volume fraction $\eta_\mathrm{Hard}(\phi_\mathrm{eff})$, with $\phi_\mathrm{eff}$ given in Eq.~\eqref{eq:phi_eff}.
In inset, an alternative form of the critical scaling, Eq.~\eqref{eq:critical:scaling:eta}, $\dot\gamma \eta_\mathrm{s} = g_\mathrm{\eta}(\Delta\phi/\dot\gamma)$, with $g_\eta(x) = f_\eta(1/x)/x$.}
\label{fig:phi_eff}
\end{figure}

The zero-shear viscosity $\mathrm{lim}_{\dot\gamma\to 0} \eta(\dot\gamma, \phi)$ is also the viscosity in the hard-sphere limit $\eta_\mathrm{Hard}(\phi)$, as we recall that with our non-dimensionalization $\dot\gamma$ is nothing but the Weissenberg number.
Since $\beta \sim \Delta\phi \equiv \phi_{\mathrm{J}}-\phi$ for small $\Delta\phi$, the viscosity diverges at the jamming transition as $\eta \sim \Delta\phi^{-1}$, to be contrasted with the stronger divergence $\eta \sim \Delta\phi^{-\nu}$ with $\nu\approx 2-2.5$, as reported in the literature (see ~\cite{dennRheologyNonBrownianSuspensions2014,guazzelliRheologyDenseGranular2018,nessPhysicsDenseSuspensions2022} and references therein).
As shown in solid lines in Fig.~\ref{fig:phi_eff}, for finite $\dot\gamma$ the viscosity first increases with $\phi$ but stays below the hard sphere viscosity, in agreement with observations, e.g.~\cite{olsson_herschel-bulkley_2012,vagbergUniversalityJammingCriticality2014,rostiRheologySuspensionsViscoelastic2018}.
Further increasing $\phi$, the viscosity reaches a maximum below jamming, and decreases close to $\phi_\mathrm{J}$, where it vanishes (dotted lines in Fig.~\ref{fig:phi_eff}).
The decrease of viscosity occurs for large $\dot\gamma/\Delta\phi$ values, for which the deformation $q$ is large, and therefore lies beyond the limit of validity of the model.

Remarkably, thanks to the relation between particle pressure and stress anisotropy, Eq.~\eqref{eq:pressure:sigma}, the particle pressure is also linear in $\dot\gamma$ at leading order, a key feature of the rheology of suspensions of hard particles (which in particular implies a finite macroscopic friction coefficient $\mu$, as we will see later). This is in contrast to polymeric systems, for which normal stresses are quadratic in $\dot\gamma$~\cite{macoskoRheologyPrinciplesMeasurements1994}. 
In our model the physical origin of this behavior is transparent: it is a direct consequence of the finite range of the repulsive force, which implies that only the compressed part of the microstructure contributes to both pressure and deviatoric stress. For a polymeric system, contributions from the compressed and elongated parts of the microstructure add up to the deviatoric stress, but cancel out for the pressure at the lowest order in deformation.
However, our model is again like polymeric visco-elastic models when it comes to the normal stress difference $N_1 =\Sigma_{11} - \Sigma_{22}$, which is quadratic in $\dot\gamma$ at leading order: in Eq.~\eqref{eq:first_order_stress}, there is no normal stress difference contribution in simple shear, but in Eq.~\eqref{eq:second_order_stress}, the second term, involving the vorticity, gives a finite contribution to $N_1$.

Within the set of approximations performed here, we find that $(\tilde{\lambda} + 2\xi)/\beta > 0$, which causes the shear thinning.
More importantly, Eq.~\eqref{eq:shear_thinning_expansion} implies that the viscosity change $\Delta \eta(\dot\gamma) = \eta(\dot\gamma) - \eta(0)$ is linear, $\Delta\eta \propto \dot\gamma^y$ with $y=1$.
This is unusual for constitutive models, at least for simple shear flows, for which other visco-elastic models in the literature give $y=2$ (for instance Johnson and Segalman~\cite{johnsonModelViscoelasticFluid1977}, Giesekus~\cite{giesekus1966elasticity,germannEnglishTranslationGiesekus2022,giesekusSimpleConstitutiveEquation1982}, Larson~\cite{larsonConstitutiveEquationPolymer1984} or Phan-Thien and Tanner~\cite{phan-thienNewConstitutiveEquation1977,phan-thienNonlinearNetworkViscoelastic1978}).
Indeed, in our model, the terms leading to $y=1$ are the ``singular'' ones involving $|\bm{\Sigma}'|$ in Eq.~\eqref{eq:sigmaprim2}.
Measured values of $y$ reported in the literature are rather diverse, but usually are $\lesssim 1$. 
Numerical simulations by Vågberg et al. show $y\approx 0.93$~\cite{vagbergUniversalityJammingCriticality2014}, while others by Kawasaki et al. show rather $y\approx 0.56$~\cite{kawasaki_diverging_2015}.
Experiments on emulsions, foams and microgels are compatible with $y\approx 0.4-0.5$~\cite{paredes_rheology_2013,dinkgreve_universal_2015}.

\begin{figure}
\includegraphics[width=0.6\columnwidth]{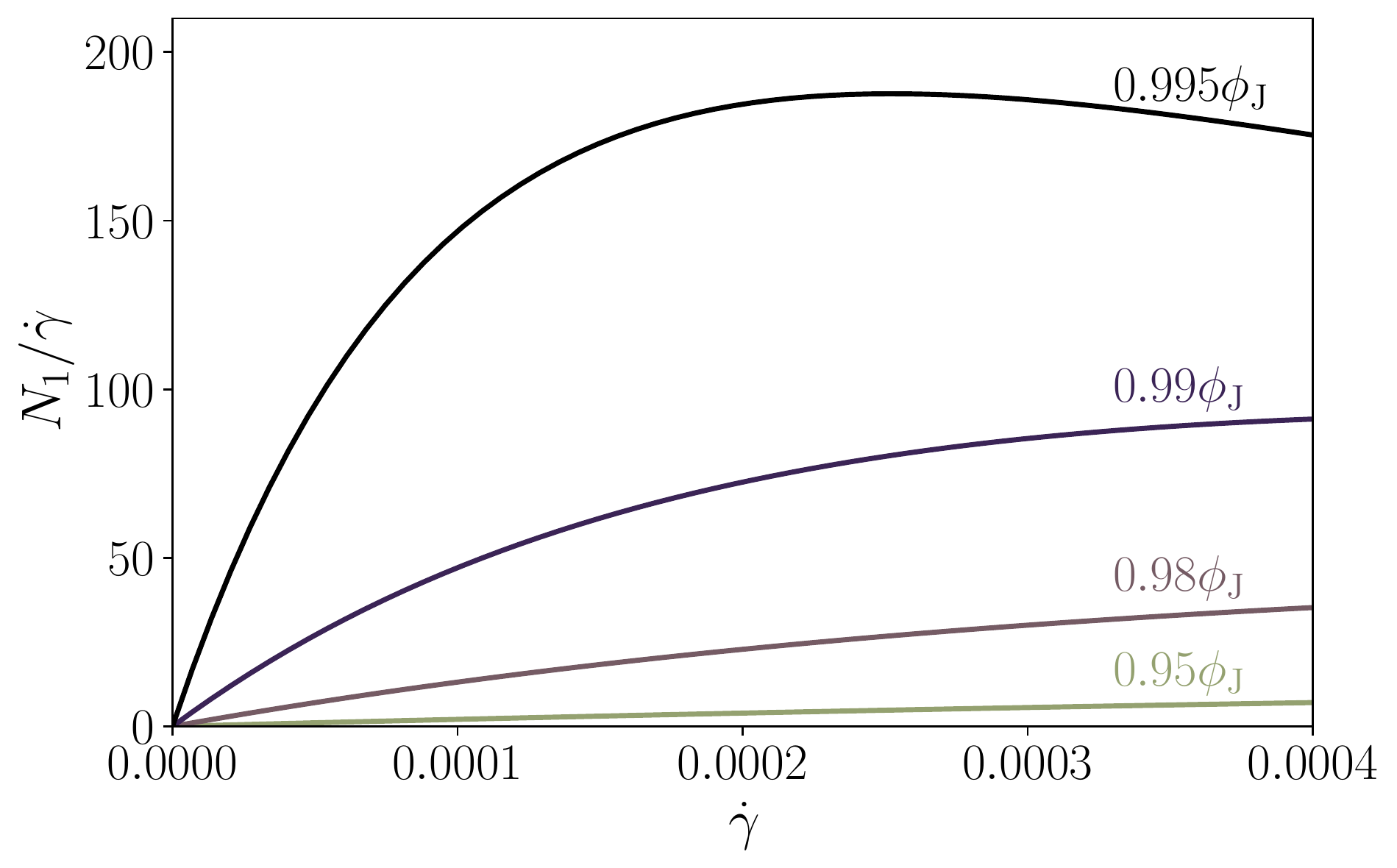}
\caption{Dimensionless first normal stress difference $N_1$ as a function of dimensionless shear rate $\dot\gamma$ for several volume fractions below the jamming volume fraction $\phi_\mathrm{J}$, under simple shear flow.}
\label{fig:N1_curves}
\end{figure}

In Fig.~\ref{fig:N1_curves}, we show the normal stress difference viscosity $N_1/\dot\gamma = (\Sigma_{11} - \Sigma_{22})/\dot\gamma$ we measure in simple shear flow, as a function of $\dot\gamma$. As expected from Eq.~\eqref{eq:second_order_stress}, $N_1$ vanishes in the small $\dot\gamma$ limit. 
For finite $\dot\gamma$, it is a concave function, which is consistent with the behavior observed in simulations of emulsions~\cite{zinchenkoShearFlowHighly2002,zinchenkoExtensionalShearFlows2015,aouaneStructureRheologySuspensions2021}.
The concavity is increasing with increasing $\phi$, and at large $\phi$ we see a decrease of $N_1/\dot\gamma$ at large $\dot\gamma$.
Such a decrease has been observed in emulsions for large viscosity ratios between the two phases~\cite{zinchenkoExtensionalShearFlows2015}.

\subsection{Critical scaling}

The question of a critical scaling for the rheology of suspensions of soft particles close to the jamming transition is recurring in the literature~\cite{olsson_critical_2007,olsson_critical_2011,paredes_rheology_2013,dinkgreve_universal_2015,kawasaki_diverging_2015}.
Close to the jamming transition, it has been proposed from experimental observations that the viscosity follows a scaling law
\begin{equation} \label{eq:critical:scaling:eta}
    \eta \Delta \phi^a = f_\eta(\dot\gamma/\Delta\phi^b),
\end{equation}
with $a\approx 1.7$ and $b\approx 3.8$~\cite{paredes_rheology_2013}. 
Numerical simulations of idealized suspensions reported a similar scaling form, but with exponents $a \approx 1.65$ and $b\approx 2.85$~\cite{olsson_critical_2007}.
Corrections to scaling are however known to be significant in the range of $\Delta\phi$ accessible in practice both in experiments and numerics, and therefore the precise evaluation of the true critical exponents is challenging~\cite{olsson_critical_2011,kawasaki_diverging_2015}.

In our model, as we have $\beta \sim \Delta\phi$, one can rewrite Eq.~\eqref{eq:shear_thinning_expansion} under the scaling form (\ref{eq:critical:scaling:eta}) with $a=b=1$, where $f_{\eta}$ is a known scaling function, independent of both $\dot\gamma$ and $\Delta\phi$. 
This critical scaling property is illustrated in the inset of Fig.~\ref{fig:flow_curves}a, for the viscosity $\eta_\mathrm{s}$ in simple shear. 
For small enough $\Delta \phi$ values, the rescaled viscosity $\eta_\mathrm{s} \Delta\phi$ falls onto a master curve as function of $\dot\gamma/\Delta\phi$, whereas we can see deviations from this master curve for $\Delta\phi = 0.05$.
Similarly, in the inset of Fig.~\ref{fig:flow_curves}b, we show the Trouton ratio $\eta_\mathrm{e}/\eta_\mathrm{s}$, which close to jamming is a function of $\dot\gamma/\Delta\phi$ only.
Alternatively, this critical scaling can be used to rescale data obtained for varying $\phi$ at fixed values of $\dot\gamma$, as in Fig.~\ref{fig:phi_eff}. 
Indeed, we have $\eta_\mathrm{s} \dot\gamma = g_\eta(\Delta\phi/\dot\gamma)$ with $g_\eta(x) = f_\eta(1/x)/x$. This alternative form is shown in the inset of Fig.~\ref{fig:phi_eff}.

One may wonder whether the critical scaling property is limited in our model to the expansion to order $\dot\gamma^2$ of the deviatoric stress tensor $\gSigma'_\mathrm{st}$ performed to derive 
Eq.~\eqref{eq:shear_thinning_expansion}, with possible corrections to scaling when taking into account higher orders in $\dot\gamma$.
We have checked that the critical scaling property remains valid when performing an expansion to order $\dot\gamma^3$ of $\gSigma'_\mathrm{st}$, corresponding to an expansion to order $\dot\gamma^2$ of the viscosity $\eta$.
This critical scaling property can be understood as follows from the steady-state version of Eq.~\eqref{eq:sigmaprim2}. Neglecting the $\Delta\phi$-dependence of other coefficients than $\beta \propto \Delta\phi$ (an assumption valid in the critical regime $\Delta\phi\ll 1$) and dividing Eq.~\eqref{eq:sigmaprim2} by $\Delta\phi$, one finds that the $\Delta\phi$-dependence can be reabsorbed into a scaled tensor $\bm{E}/\Delta\phi$. The only dependence on $\dot\gamma$ that is not rescaled by $\Delta\phi$ is in the vorticity $\bm{\Omega}$. 
Recalling that the pressure $p$ appearing in the term $2p\bm{E}$ in Eq.~\eqref{eq:sigmaprim2} is proportional to $|\gSigma'|$ from Eq.~\eqref{eq:pressure:sigma}, one finds that the rescaled viscosity $\eta \Delta\phi = (\bm{\Sigma}'_\mathrm{st}:\hat{\bm{E}})\Delta\phi/\dot\gamma$ is a function of $\dot\gamma/\Delta\phi$ as long as the projection $\bm{\Sigma}'_\mathrm{st}:\hat{\bm{E}}$ of the deviatoric stress tensor $\bm{\Sigma}'_\mathrm{st}$ on the normalized strain rate tensor $\hat{\bm{E}}$ is independent of the vorticity ${\bm\Omega}$. This property remains true at least up to order $\dot\gamma^3$ in the expansion of $\bm{\Sigma}'_\mathrm{st}$, as mentioned above, thanks to the relation $(\hat{\bm{\Omega}}\cdot \hat{\bm{E}} - \hat{\bm{E}}\cdot \hat{\bm{\Omega}}):\hat{\bm{E}} = 0$. Whether it is valid at all orders in $\dot\gamma$ is a difficult question, which we do not attempt to address here. In any case, Eq.~\eqref{eq:sigmaprim2} is expected to be valid only at low shear rate due to the weakly anisotropic expansion performed in Sec.~\ref{sec:stress-tensor}, and considering a high-order expansion of Eq.~\eqref{eq:sigmaprim2} in $\dot\gamma$ may not be physically relevant.


Interestingly, the critical scaling (\ref{eq:critical:scaling:eta}) of the viscosity also sheds light on the idea mentioned in the introduction that particles effectively appear softer when increasing shear stress. The viscosity of soft particles may then be approximated as $\eta_\mathrm{Hard}(\phi_\mathrm{eff})$, where $\eta_\mathrm{Hard}(\phi)$ is the viscosity for a suspension of hard spheres at volume fraction $\phi$~\cite{trulsson_athermal_2015}.
As $\eta_\mathrm{Hard}(\phi)$ corresponds in our model to the limit of $\eta_{\mathrm{s}}(\dot\gamma,\phi)$ when the dimensionless shear rate $\dot\gamma$ goes to zero, one can thus define $\phi_\mathrm{eff}$ through the relation $\eta_{\mathrm{s}}(\dot\gamma,\phi)=\eta_{\mathrm{s}}(0,\phi_\mathrm{eff})$.
Using the critical scaling given in Eq.~(\ref{eq:critical:scaling:eta}), one finds
\begin{equation}
    \phi_\mathrm{eff} = \phi_\mathrm{J} - \Delta\phi \frac{f_\eta(0)}{f_\eta(\dot\gamma/\Delta\phi)}\,.
\end{equation}
At small shear rate $\dot\gamma$, $\phi_\mathrm{eff}$ is close to the nominal volume fraction $\phi$, and it decreases with increasing shear rate (since $f_{\eta}$ is a decreasing function, see inset of Fig.~\ref{fig:flow_curves}a), in agreement with physical intuition.
For small enough $\dot\gamma/\Delta\phi$, one finds
\begin{equation}
    \phi_\mathrm{eff} = \phi - \frac{\tilde{\lambda}+2\xi}{2\beta_0} \phi_\mathrm{J} \dot\gamma\,,
    \label{eq:phi_eff}
\end{equation}
so that $\phi_\mathrm{eff}-\phi$ is actually independent of $\Delta\phi$ to leading order in $\dot\gamma$.
The approximation $\eta(\phi, \dot\gamma)\approx \eta_\mathrm{Hard}(\phi_\mathrm{eff}(\dot\gamma))$ is in dashed lines in Fig.~\ref{fig:phi_eff}.


\subsection{Soft granular rheology}

\begin{figure}
\includegraphics[width=0.5\columnwidth]{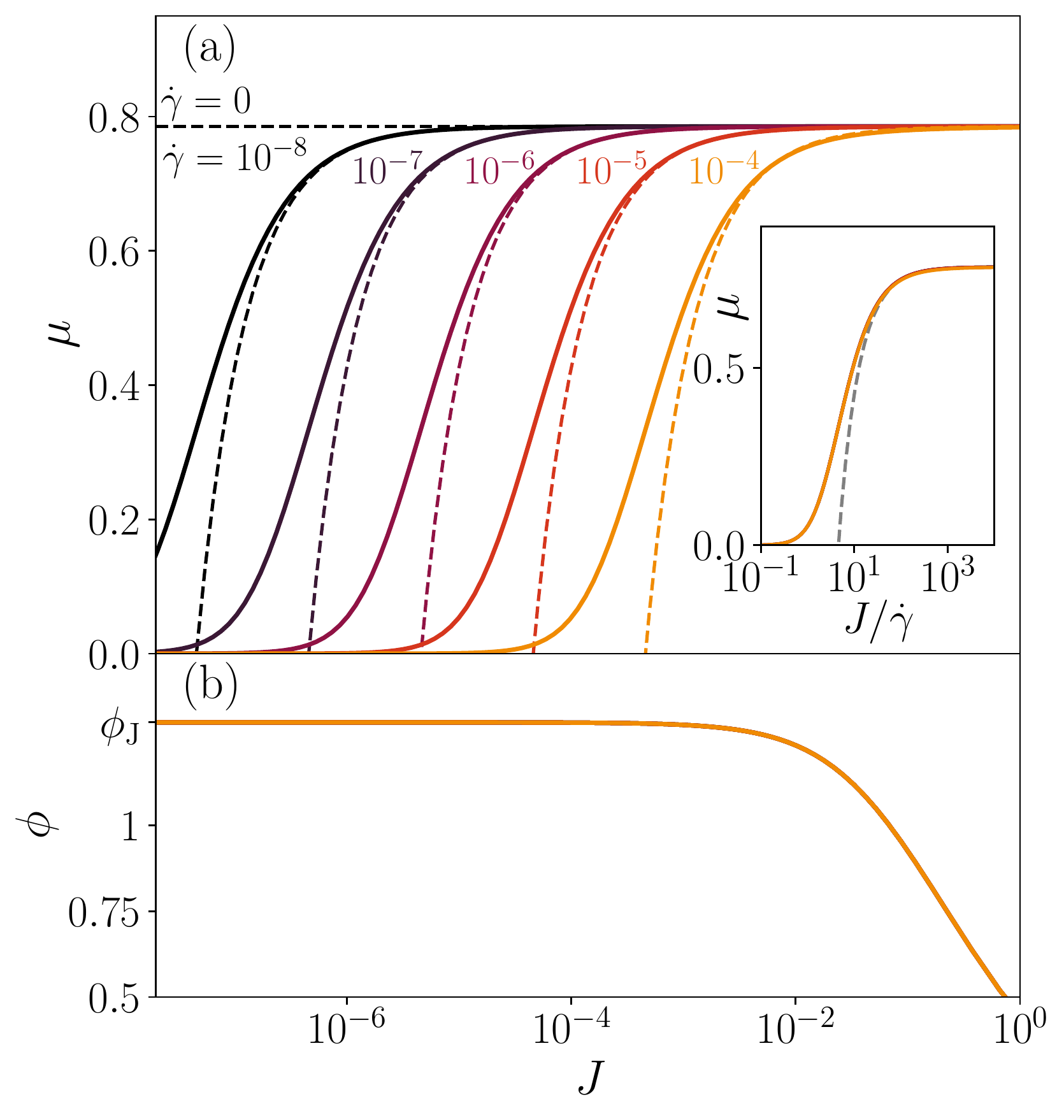}
\caption{(a) Macroscopic friction coefficient $\mu = \Sigma_{12}/p$ as a function of the viscous number $J = \dot\gamma/p$ in simple shear, for several values of the dimensionless shear rate (Weissenberg number) $\dot\gamma$, in solid lines. In dashed lines, predictions from  Eq.~\eqref{eq:mu_J_dotgamma}. In inset the same data plotted as a function of $J/\dot\gamma$, showing that all shear rates collapse on the same curve, which is predicted to be exact at the lowest order in $J$ and $p$ by Eq.~\eqref{eq:mu_J_dotgamma} (prediction shown in dashed line). (b) Volume fraction as a function of $J$, for the same values of the dimensionless shear rate $\dot\gamma$. The effect of softness is not visible, as it scales as $\dot\gamma$, which largest value here is $\dot\gamma = \num{e-4}$. }
\label{fig:mu_J_Ca}
\end{figure}


In the previous section, we presented the rheology under constant volume, as our control parameter was the packing fraction $\phi$.
Alternatively, the rheology can be expressed in the framework of constant particle pressure rheology. 
In this framework, for soft particles one introduces dimensionless numbers $\mu$, $J$ and $p$ to characterize the simple shear flow of a dense suspension, with
\begin{equation}
\mu = \frac{\Sigma_{12}}{p}\,, \quad J = \frac{\gd}{p}\, .
\end{equation}
(We recall that $\dot\gamma$ and $p$ are made dimensionless, see Sec.~\ref{sec:model}.)
Scalar constitutive equations for the shear components of the stress and strain rate tensors can then be formulated in terms of the two functions $\mu(J, p)$ and $\phi(J, p)$.
One may alternatively use the dimensionless shear rate $\dot\gamma$ instead of the dimensionless pressure $p$, in which case the rheology is expressed as  $\mu(J,\dot\gamma)$ and $\phi(J,\dot\gamma)$.
Sticking to $J$ and $p$ as dimensionless numbers,
for low particle softness, corrections to the jamming limit $J,p\to 0$ have been discussed in \cite{kawasaki_diverging_2015}, and take the generic form
\begin{align}
    \label{eq:muJP:gen}
   \mu(J,p) &= \mu_{\rm J}  + b_{\mu} J^{\beta_\mu} - c_{\mu} p^{\alpha_\mu},\\
   \label{eq:phiJP:gen}
   \phi(J,p) &= \phi_{\rm J}  - b_{\phi} J^{\beta_\phi}+ c_{\phi} p^{\alpha_\phi}.
\end{align}
Exponent values $\beta_\mu \approx 0.346$, $\beta_\phi\approx 0.391$, $\alpha_\mu \approx 0.56$ and $\alpha_\phi \approx 0.75$ have been evaluated in numerical simulations of a three-dimensional suspension of harmonic spheres~\cite{kawasaki_diverging_2015}.


This soft granular rheology approach can also be applied to our constitutive model.
The tensorial constitutive equation \eqref{eq:sigmaprim2} allows us to generalize the soft granular rheology relation $\mu(J,p)$ into a tensorial form, by introducing tensors ${\bm\mu}=\gSigma'/p$ and ${\bm J}=\bm{E}/p$. At lowest order, we find
\begin{equation} \label{eq:muJ:tensorial}
    {\bm\mu} = {\bm\mu}_{\rm J} - \frac{\pi^2}{8\kappa} (\tilde{\lambda}+2\xi) p \hat{\bm E}
    + \frac{\pi}{\kappa} p [\hat{\bm{\Omega}} \gdot \bm{J} -\bm{J} \gdot \hat{\bm{\Omega}} ]
\end{equation}
with ${\bm\mu}_{\rm J} = \frac{\pi}{2} \hat{\bm{E}}$. 
Several comments are in order. First, the fact that our theory gives a nonzero value of ${\bm \mu}_{\rm J}$ is already a nontrivial result, and is again a consequence of the finite range of the repulsive interaction in our model.
Second, the vorticity potentially brings a non-trivial tensorial contribution to ${\bm \mu}$, in such a way that the tensor ${\bm\mu}$ is not necessarily proportional to $\hat{\bm E}$.
%
%

However, in the case of a simple shear flow geometry, the effect of vorticity on the shear component of Eq.~\eqref{eq:muJ:tensorial} disappears.
We define the off-diagonal components of 
${\bm\mu}$ and ${\bm J}$ as $\mu$ and $J/2$ respectively, to match standard definitions \cite{kawasaki_diverging_2015,favierdecoulombRheologyGranularFlows2017}.
We get for the lowest order expansions of $\mu(J,p)$ and $\phi(J,p)$ the simple form
\begin{equation}
   \mu(J,p) = \mu_{\rm J} - c_{\mu} p\,, \qquad
   \phi(J,p) = \phi_{\rm J} - b_{\phi} J\,,\label{eq:mu_J_p_lowest_order}
\end{equation}
with $\mu_{\rm J}=\pi/4$.
By comparison with Eq.~\eqref{eq:muJP:gen}, we find $\alpha_\mu=\beta_\phi=1$ and $b_{\mu}=c_{\phi}=0$ while the other coefficients $c_{\mu}$ and $b_{\phi}$ are nonzero:
\begin{equation}
c_{\mu} = \frac{\pi^2}{16\kappa} (\tilde{\lambda}+2\xi)\,, \qquad b_{\phi} = \frac{2\kappa\phi_\mathrm{J}}{\pi \beta_0} \,.
\end{equation}
Note that the term involving the vorticity disappears as it has no off-diagonal component in a simple shear flow.

The decrease of $\mu$ with increasing $p$ is well documented for soft particles~\cite{kawasaki_diverging_2015,favierdecoulombRheologyGranularFlows2017}, or even for Brownian particles~\cite{wangConstantStressPressure2015,billonTransitionGranularBrownian2022}, as Brownian motion has somewhat similar effects on rheology as softness~\cite{trulsson_athermal_2015}.
The reason why no correction in $J$ appears in the expansion of $\mu(J,p)$ for our model can be traced back to the very simple parametrization of the anisotropic pair correlation function introduced in Eq.~\eqref{param_g}, which forces the deviatoric stress tensor to be proportional to the pressure with a proportionality factor independent of $\Delta\phi$. On the other side, the reason for the absence of a correction in $p$ in the expansion of $\phi(J,p)$ lies in the low order of the expansion. Going to the next order in the expansion of $\phi(J,p)$ requires to perform the expansion of the pressure to order $q^2$, thus extending Eq.~\eqref{eq:pressure:q} to the next order. In this way, one finds the first correction of $\phi(J,p)$ in $p$, which takes the form
\begin{equation}
   \phi(J,p) = \phi_{\rm J} - b_{\phi} J + c_{\phi}' Jp
   \label{eq:phi_J_p}
\end{equation}
(where $c_{\phi}'$ is a known coefficient), meaning that the relevant variables to evaluate the corrections of $\phi$ around $\phi_{\rm J}$ are rather $J$ and $\dot\gamma = Jp$.

In Fig.~\ref{fig:mu_J_Ca}a, we plot the results of numerical integration of Eq.~(\ref{eq:sigmaprim2}) for $\mu$. 
As control parameters in Eq.~(\ref{eq:sigmaprim2}) are $\phi$ and $\dot\gamma$, while $\mu$ and $J$ are outcomes, we present results as $\mu$ as a function of $J$ for several fixed (small) values of the dimensionless shear rate $\dot\gamma$. 
To evaluate the pressure we used Eq.~\eqref{eq:pressure:sigma} with $|\gSigma'|$ evaluated at first order in $\dot\gamma$ [Eq.~\eqref{eq:first_order_stress}] to remain consistent in our expansion in $q$.
The predictions of the lowest order expansion in $p=\dot{\gamma}/J$, obtained from Eq.~\eqref{eq:mu_J_p_lowest_order} as
\begin{equation}
   \mu(J,\dot{\gamma}) = \mu_{\rm J} - c_{\mu}\, \frac{\dot{\gamma}}{J}\,,
   \label{eq:mu_J_dotgamma}
\end{equation}
are shown in dashed line for comparison. We recover a plateau of value $\mu=\mu_\mathrm{J}$ at large $J$, corresponding to a small pressure $p$.
At low $J$ values (keeping $\dot{\gamma}$ fixed), particle softness induces a decrease in $\mu$, and we find that in the $J\to 0$ limit, $\mu$ vanishes for all finite $\dot \gamma$. 
It should however be noted that in this limit the microstructure anisotropy $q$ is large, and the model reaches its limits of validity.

According to Eq.~\eqref{eq:mu_J_p_lowest_order}, the curves for several values of $\dot\gamma$ can be collapsed by plotting them as a function of $J/\dot\gamma$, as long as $p=\dot\gamma/J$ is small enough. This rescaling is plotted in the inset of Fig.~\ref{fig:mu_J_Ca}, which shows the quality of the collapse even for surprisingly large values of $p$ of order 1.

In Fig.~\ref{fig:mu_J_Ca}b, we show the corresponding numerical results for $\phi$ as function of $J$ for the same dimensionless shear rate values as in Fig.~\ref{fig:mu_J_Ca}a. 
Following Eq.~\eqref{eq:phi_J_p}, the effect of varying the shear rate is of order $\dot\gamma$, and is indeed unnoticeable in Fig.~\ref{fig:mu_J_Ca}b.

\subsection{Transients}

\begin{figure}
\includegraphics[width=0.55\columnwidth]{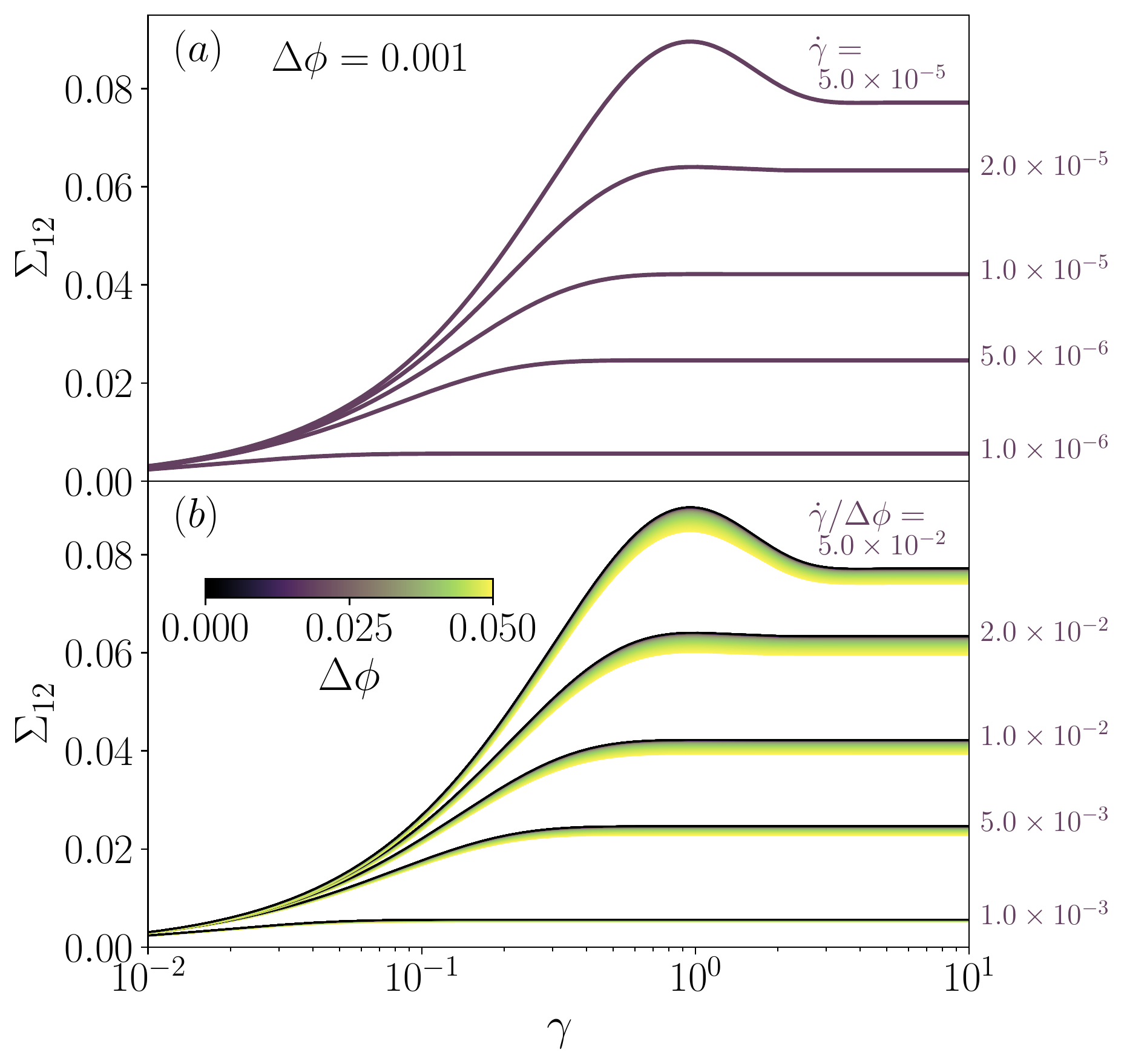}
\caption{(a) Load curve in simple shear, $\Sigma_{12}$ as a function of strain $\gamma$, for $\Delta\phi=0.001$ and several values of shear rate $\dot{\gamma}$, increasing from bottom to top. (b) Load curves for several $\Delta\phi$ values (increasing from dark to light color) and several $\dot{\gamma}$, bundled by groups sharing the same scaled shear rate values $\dot\gamma/\Delta\phi$, increasing from bottom to top. Load curves collapse on a master curve in the small shear rate limit, showing the critical scaling of the dynamics at work.}
\label{fig:transients}
\end{figure}

We now investigate the transient rheology predicted by our model. 
In Fig.~\ref{fig:transients}a, we show the predicted load curves in simple shear starting from an initial resting condition ${\bm \Sigma}' = 0$, for several values of the shear rate. 
At small shear rates, load curves are monotonic and steady state is reached after a strain of order $1$. 
At rates $\dot\gamma \gtrsim 10^{-5}$, steady state takes longer to achieve, up to a strain $\gamma\approx 5$, and the suspension passes through a stress overshoot for $\gamma \approx 1$. 
There are few reports of experimental load curves for suspensions of soft particles below jamming, but stress overshoots have been observed in polymer blend emulsions~\cite{bousminaRheologyPolymerBlends2001a}, or in simulations of emulsions~\cite{loewenbergNumericalSimulationConcentrated1996}.

Remarkably, in our model a scaling form also holds for the temporal evolution of the stress. Indeed, following the same reasoning that concludes to the existence of scaling in steady state, we find that asymptotically close to jamming, the stress follows ${\bm \Sigma}' = f_{\Sigma'}(\dot\gamma/\Delta\phi, t\Delta\phi)$. 
This implies that stress-strain load curves at different $\Delta\phi$ can be superimposed if compared for the same values of $\dot\gamma/\Delta\phi$.
This is done in Fig.~\ref{fig:transients}b, showing the presence of scaling as well as deviations from it when $\Delta\phi$ exceeds a few percents.



\begin{figure}
\includegraphics[width=0.5\columnwidth]{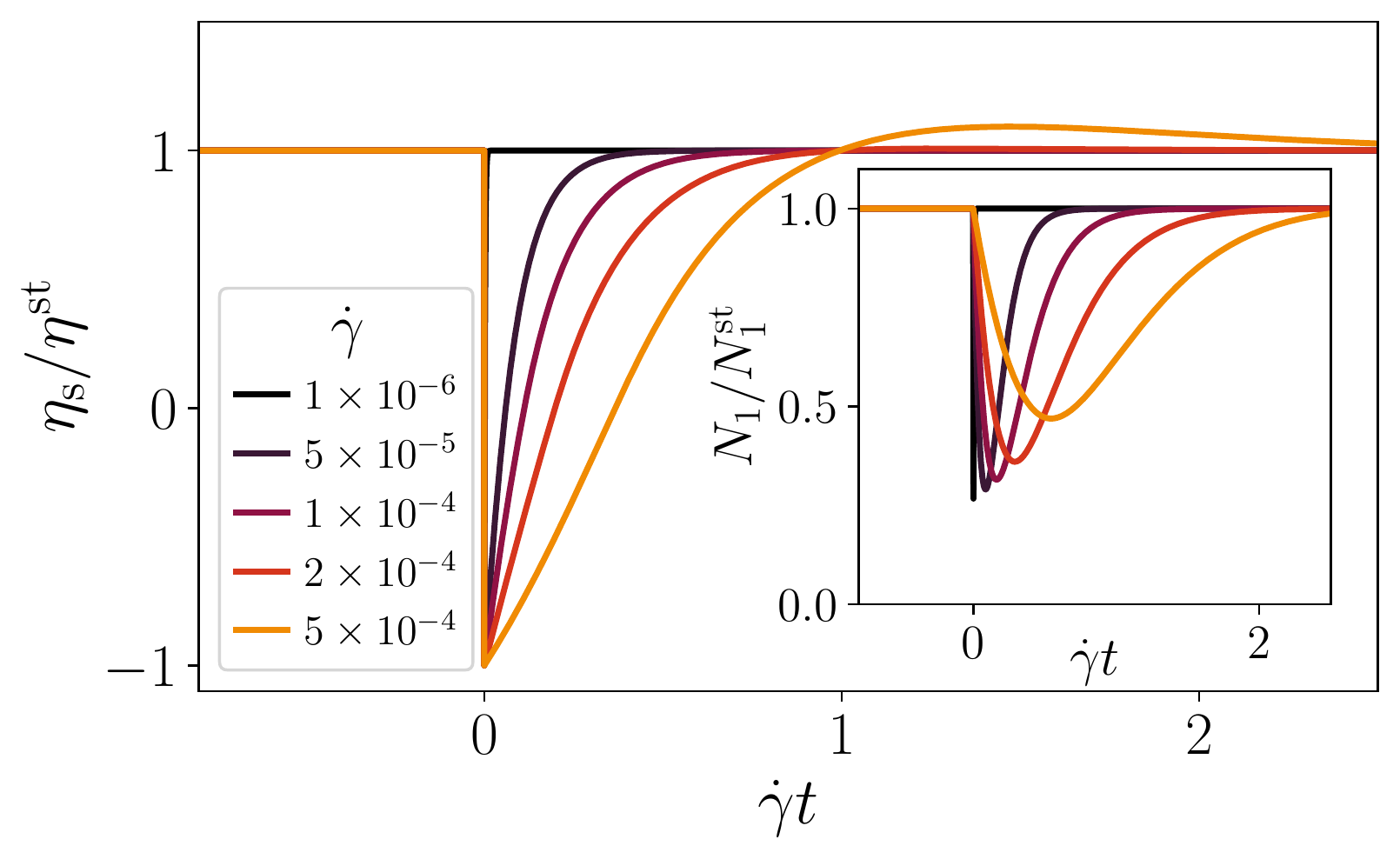}
\caption{Shear reversal at $\Delta\phi = 0.01$ for several values of shear rate $\dot\gamma$, increasing from dark to light colors. In the main panel, the viscosity $\eta_\mathrm{s}$ scaled by its steady-state value $\eta^\mathrm{st}$ as a function of strain after reversal. In inset, the scaled normal stress difference $N_1/N_1^\mathrm{st}$.}
\label{fig:reversal}
\end{figure}

We also investigate the predictions of the model in shear reversal, where starting from a steady-state simple shear flow under shear rate $\dot\gamma$, one suddenly reverses the flow direction $\nabla \bm{u} \to -\nabla \bm{u}$.
The viscosity and normal stress difference $N_1 =\Sigma_{11} - \Sigma_{22}$ scaled by their respective steady-state values $\eta^\mathrm{st}$ and $N_1^\mathrm{st}$ are shown as a function of post-reversal strain in Fig.~\ref{fig:reversal} and its inset.
For all shear rates, $\eta_\mathrm{s}/\mathrm{\eta}^\mathrm{st}$ at reversal discontinuously jumps from $1$ to $-1$, reflecting the fact that the configuration did not change at reversal.
For the same reason, $N_1/N_1^\mathrm{st}$ is continuous at reversal.
The later evolution towards the steady-state value $\eta_\mathrm{s}/\mathrm{\eta}^\mathrm{st}=1$ depends on the shear rate.
At large shear rates, the viscosity has a non-monotonic evolution to steady state, with an overshoot at  strains of order $1$. 
The normal stress difference first decreases before increasing back up to its steady-state value.
By contrast, at low rates, the relaxation is monotonic and very quick. 
In fact, in the limit of vanishing shear rates (i.e. the hard particle limit), the viscosity (and normal stress difference) returns to steady state after a vanishing strain. 
This behavior is of course quite different from what is observed for suspensions of hard particles~\cite{gadala-mariaShearInducedStructure1980,narumiTransientResponseConcentrated2002, kolliTransientNormalStress2002,blancLocalTransientRheological2011}, for which the post-reversal transients extend over a finite strain, and are typically non-monotonic as $\eta_\mathrm{s}/\mathrm{\eta}^\mathrm{st}$ passes through a minimum.

This behavior exposes limits of our model coming from the simplified treatment of the microstructure which is reduced to the anisotropy of the contact shell. 
In the hard-particle limit the dynamics of contacts, being driven by elastic forces, is infinitely fast, and thus our model predicts no shear reversal transients. 
However, in an actual suspension part of the stress also comes from particle in near (but not quite in) contact, which anisotropic structure evolves with strain, and not only elastic forces.
This stress contribution can in some cases be neglected, as in steady state, where it is much smaller than the contributions from contacts close to jamming~\cite{mariShearThickeningFrictionless2014,gallierRheologyShearedSuspensions2014,linHydrodynamicContactContributions2015}. 
In shear reversal however, the contact contribution is transiently strongly suppressed, and taking into account other stress sources becomes essential for an accurate prediction of the transient~\cite{petersRheologyNonBrownianSuspensions2016}.

\section{Discussion and conclusion}

We developed a constitutive model to describe the flow behavior of dense non-Brownian suspensions of soft elastic disks.
Starting from the microscopic equations of motion, we could express the dynamics of the stress as an ensemble average of the time evolution of the pair correlation function.
The resulting constitutive model, Eq.~\eqref{eq:sigmaprim2}, describes a visco-elastic shear-thinning behavior, with a zero-shear-rate viscosity $\eta(\dot\gamma\to 0)$ diverging at the jamming transition. 

However our model has several unique and appealing features. 
First, while the normal stress difference is quadratic in shear rate at leading order, $N_1\propto \dot\gamma^2$, the particle pressure has a linear contribution in the shear rate $\dot\gamma$, which eludes visco-elastic models known in the literature. 
This contribution is a direct consequence of the finite range of contact forces, which implies that pairs of particles in the compressed and elongated quadrants of the microstructure do not have opposite contributions to the isotropic part of normal stresses, and leads to Eq.~\eqref{eq:pressure:sigma}.
These distinct scalings for normal stresses and their differences is a distinguishing feature of emulsions~\cite{zinchenkoShearFlowHighly2002,zinchenkoExtensionalShearFlows2015,malekiViscousResuspensionDroplets2022}.

Second, the shear-thinning behavior is uncommon among models of visco-elastic materials, which usually predict $\eta(\dot\gamma)-\eta(0)\propto \dot\gamma^2$.
By contrast, our model predicts $\eta(\dot\gamma)-\eta(0)\propto \dot\gamma$, while available observations for suspensions of soft particles report $\eta(\dot\gamma)-\eta(0)\propto \dot\gamma^y$ with $y$ ranging from $\approx 0.4$ to $\approx 1$, depending on the authors~\cite{loewenbergNumericalSimulationConcentrated1998,zinchenkoShearFlowHighly2002,zinchenkoExtensionalShearFlows2015,zinchenkoGeneralRheologyHighly2017,paredes_rheology_2013,dinkgreve_universal_2015,vagbergUniversalityJammingCriticality2014,kawasaki_diverging_2015}.
This unusual prediction stems from the presence of ``singular'' terms involving $|\bm{\Sigma}'|$, which do not naturally arise when developing phenomenological models following Hand theory, as $|\bm{\Sigma}'|$ is the square-root of a tensorial invariant.
The general approach developed here shares its starting point, a formally exact but unclosed stress evolution from the Smoluchowski equation for the dynamics of the pair correlation function, with the standard method used to derive constitutive models of polymer solutions or melts (sometimes called ``Smoluchowski'' theory).
Usual polymeric models derived from microscopics however do not contain similar singular terms.
Indeed, in our model these terms can be traced back to the fact that extensional quadrants around a given particle do not contribute to the tensorial integrals \eqref{def_Theta} and \eqref{def_Phi} that describe the interplay of the particle-pair dynamics with the applied flow, while extensional quadrants contribute to the stress evolution in polymeric systems.
A further singular term involving $|\bm{\Sigma}'|\bm{\Sigma}'$ appears in our approach if we push the expansion to the next order in the anisotropy parameter $q$. However it turns out this term comes with a prefactor turning the rheology to shear thickening. This calls for improvements in the closures we use in our method.

While many technical challenges faced here, such as closures, are not specific to the softness of the particles, the constitutive model is not suitable for the description of suspensions of hard particles. 
This is obvious when looking at the dynamics under shear reversal. In the limit of infinitely stiff particles, our model predicts that at reversal the stress jumps instantaneously from its steady-state value in the forward direction to its steady-state value in the backward direction. Our model is thus oblivious to the transient decrease of viscosity observed 
on a strain of order one.
Besides, our model successfully captures the fact that suspensions of hard particles have a finite macroscopic friction coefficient $\mu$ at jamming, in the limit of vanishing viscous number, $J \to 0$.
However, $\mu$ is predicted as independent of $J$ in the limit of small dimensionless shear rate (i.e., Weissenberg or capillary number), whereas experiments and simulations show that $\mu$ is an increasing function of $J$. 
These deficiencies are tied to assumptions about the first-neighbor shell. First, the number of particles in the shell (measured by the weight $A$) has no intrinsic dynamics, and second, the distribution of particles within the shell remains homogeneous (even if the shell distorts, and creates an anisotropic microstructure). 
Both these assumptions could probably be relaxed in an extension of the presented model.
It is however quite uncertain at this stage whether such an extended model would naturally bridge to established models for hard sphere suspensions~\cite{hinchConstitutiveEquationsSuspension1975,hinchConstitutiveEquationsSuspension1976,phan-thienConstitutiveEquationConcentrated1995,goddardDissipativeAnisotropicFluid2006,stickelConstitutiveModelMicrostructure2006}, and in particular the recent Gillissen-Wilson model~\cite{gillissenModelingSphereSuspension2018,gillissenConstitutiveModelTimeDependent2019,gillissenConstitutiveModelShearthickening2020}.

Such a bridge model should also be able to recover and enlighten the scaling crossover that is expected for the normal stress differences as a function of $\dot\gamma$. Indeed, in the limit of small shear rates, one should recover the hard sphere scaling $N_1, N_2 \sim \dot\gamma$~\cite{dennRheologyNonBrownianSuspensions2014,guazzelliRheologyDenseGranular2018,nessPhysicsDenseSuspensions2022}, which should crossover at finite $\dot\gamma$ to $N_1, N_2 \sim \dot\gamma^2$~\cite{loewenbergNumericalSimulationConcentrated1996,zinchenkoShearFlowHighly2002,zinchenkoGeneralRheologyHighly2017}. 
To our knowledge, no constitutive model is able to predict such crossover at the moment.

Finally, while our derivation gives us access to a model which coefficients are known functions of the microscopic particle properties, the closures and approximations we perform affect the coefficient values and lead to predictions that are not quantitative. 
Nonetheless, given the merits of the model on several qualitative predictions, as discussed above, it is possible that our model becomes quantitatively accurate if the several coefficients involved in Eq.~\eqref{eq:sigmaprim2} are considered as adjustable parameters of the model, with $\beta$ as a special case as it should remain a function of $\Delta\phi$ to preserve the existence of the jamming transition, where $\beta$ vanishes.


\begin{acknowledgments}
This work is supported by the French National Research Agency in the framework of the ``Investissements
d’avenir'' program (ANR-15-IDEX-02).
\end{acknowledgments}


\appendix

\section{Evaluation of tensorial integrals}
\label{app:integrals}

\subsection{Tensors defined by an integral over $\gtwo$}

We evaluate here the traceless parts $\bm{\Theta}'$, $\bm{\Phi}'$, $\bm{\Xi}'$ and $\bm{\Pi}'$ of the corresponding tensors introduced in Eqs.~\eqref{def_Theta} to \eqref{def_Pi}, which are defined by integrals over the pair correlation function $\gtwo(\bm{r})$.
We start by the computation of $\bm{\Pi}'$, which can be performed exactly, without relying
on a weakly anisotropic parametrization of $\gtwo(\bm{r})$.
With the form $f(r)=r-2$ of the interparticle force chosen here, the transposed gradient $\nabla\gF^\mathrm{T}$ reads
\begin{equation}
        \label{grad_f_linear}
        \nabla\gF^\mathrm{T} = f'(r)\, \eroer  + \frac{f(r)}{r}\, \et\otimes\et = \eroer + \frac{r-2}{r}\, \et\otimes\et,
\end{equation}
whence the relation
\begin{align}
        \label{rof_grad_f_T}
        \left(\rof\right)\gdot\nabla\gF^\mathrm{T}
       = \rof
\end{align}
follows, and we find that $\gPi=\gSigma$. Therefore we have
\begin{align}
        \label{Pip_uj}
        \gPi' & = \gSigma'.
\end{align}
In contrast, the evaluation of the other tensors $\bm{\Theta}'$, $\bm{\Phi}'$ and $\bm{\Xi}'$
in terms of the tensors $\bm{E}$ and $\gSigma'$ relies on the weakly anisotropic parametrization \eqref{param_g} of $\gtwo(\bm{r})$ and a small $q$ expansion.

The tensors $\gTheta$ and $\gPhi$ both include the strain-rate tensor $\bm{E}$ in their definition.
To proceed with the calculation, we need to use a general and explicit expression of the tensor $\bm{E}$. The latter being a symmetric traceless tensor, it can be written without loss of generality as
\begin{equation}
\bm{E}=\Msymt{a_1}{a_2}.
\end{equation}
We then have
\begin{equation}
        \label{einf_eroer}
        \bm{E}:\eroer= a_1\cos2\theta + a_2\sin2\theta.
\end{equation}
By replacing this expression in the definition of $\gTheta$ and $\gPhi$ and by applying the same calculation steps as for the evaluation of the stress tensor, Eq.~\eqref{def_sigmap}, we obtain after truncation at order $q$ of the small $q$, weakly anisotropic expansion,
\begin{align}
        \label{Thetap_uj}
        \gTheta' & = \rho^2 \left(\pi A-\frac{8}{3}(2A-1)q \right) \bm{E} -\frac{8}{3} q a_1(2A-1)\rho^2 \Msymt{1}{0}, \\ 
        \label{Phip_uj}
        \gPhi' & = -\frac{4}{3} A q\rho^2 \bm{E} -\frac{4}{3} Aqa_1\rho^2 \Msymt{1}{0}.
\end{align}
In these equations, one can recast $q$ in terms of $\gSigma'$ and its norm $|\gSigma'|$ using
Eqs.~\eqref{calcul_sigmaq} and \eqref{calcul_sigmaq_norm}.
One then has to reexpress $a_1$ in tensorial form. With this aim in mind, we note that the following tensorial double contraction is proportional to $a_1$,
\begin{equation}
        \label{sigmap_Einf}
        \gSigma':\bm{E}= 2\pi A\rho^2 q a_1.
\end{equation}
Eliminating in this way $q$ and $a_1$ from Eqs.~\eqref{Thetap_uj} and \eqref{Phip_uj}, one ends up with the final expressions of $\gTheta'$ and $\gPhi'$ given in Eqs.~\eqref{def_Theta_exp} and \eqref{def_Phi_exp} of the main text.

Turning to the evaluation of the tensor $\gXi'$, we find that its leading contribution is of order $q^2$, so that $\gXi'$ can be neglected in an expansion at order $q$.

\bigskip
\subsection{Tensors defined by an integral over $\gthree$}

We now have to calculate the tensors $\gGamma$ and $\gUpsilon$ defined as integrals of $\gthree$. To do so,
similarly to the case above the jamming density studied in \cite{CunyJSTAT22}, we use the so-called Kirkwood closure relation \eqref{eq:Kirkwood}
allowing one to approximate $\gthree$ by a function of $\gtwo$.
%
%
By replacing $\gtwo$ by its parametrization~\eqref{param_g}, $\gthree$ can then be expressed as
\begin{align}
        \label{g3_uj}
        \gthree(\R{},\rprim) &= \giso\left(\frac{r}{1+q\cos2\theta}\right)\, \giso\left(\frac{r'}{1+q\cos2\theta'}\right) \giso \left(\frac{u}{1+q\cos2\psi}\right),
\end{align}
with $r=|\R{}|$, $\theta=\arg(\R{})$, $r'=|\rprim|$, $\theta'=\arg(\rprim)$, $u=|\R{}-\rprim|$ and $\psi=\arg(\R{}-\rprim)$. These last two variables can be expressed as functions of $r$, $r'$, $\theta$ and $\theta'$ as
\begin{align}
        \label{exp_u}
        u & = \sqrt{r^2+r'^2-2rr'\cos(\theta-\theta')}, \\
        \label{exp_psi}
        \cos2\psi & = \frac{1}{u^2} \, [r^2\cos2\theta+r'^2\cos2\theta'-2rr'\cos(\theta+\theta')].
\end{align}
The detailed derivation of these relations can be found in \cite{CunyJSTAT22}.

Evaluating the tensor $\gGamma'$ in the weakly anisotropic limit ($q\ll 1$), one finds that the leading contribution is of order $q^2$.
As we are performing an expansion to order $q$, the tensor $\gGamma'$ can thus be neglected.

We now evaluate the tensor $\gUpsilon'$ to order $q$ in the small $q$ expansion. We start with the following identity,
\begin{equation}
        \label{rofDF}
        \left(\R{}\!\otimes\!\gF(\rprim)\right)\gdot\left(\nabla\gF(\R{})\right)^{T} \! = rf(r')f'(r)\cos(\theta'-\theta)\, \eroer + f(r')f(r)\sin(\theta'-\theta)\, \er\otimes\et\,.
\end{equation}
Using $f'(r)=1$ (since $f(r)=r-2$), one can decompose $\gUpsilon'$ into two contributions
$\gUpsilon' = \gUpsilon_\theta' + \gUpsilon_r'$, resulting from the two terms in the r.h.s.~of Eq.~(\ref{rofDF}). Using again symmetry arguments in the integration over $\theta$ and $\theta'$, we can express $\gUpsilon_\theta'$ and $\gUpsilon_r'$ as
%
%
\begin{align}
        \gUpsilon_\theta' &= \frac{\rho^3}{2} \int_I \D\theta \int_{\pi/4}^{3\pi/4} \D\theta' \int_0^2 \D r \, rf(r) \int_0^2 \D r' \, r'f(r') \sin(\theta-\theta') \sin(2\theta)
        \nonumber\\
        & \qquad \qquad \qquad \times\giso\left(\frac{r}{1+q\cos2\theta}\right)\giso\left(\frac{r'}{1+q\cos2\theta'}\right)\giso\left(\frac{u}{1+q\cos2\psi}\right) \; \Msymt{1}{0},
        \label{exp_upsilon_theta}\\
        \gUpsilon_r' &= \frac{\rho^3}{2} \int_I \D\theta \int_{\pi/4}^{3\pi/4} \D\theta' \int_0^2 \D r \, r^2 \int_0^2 \D r' \, r'f(r') \cos(\theta-\theta') \cos(2\theta)
        \nonumber\\
        & \qquad \qquad \qquad \times\giso\left(\frac{r}{1+q\cos2\theta}\right)\giso\left(\frac{r'}{1+q\cos2\theta'}\right)\giso\left(\frac{u}{1+q\cos2\psi}\right) \; \Msymt{1}{0}, \label{exp_upsilon_r}
\end{align}
where $I=[-3\pi/4,\pi/4]\cup [\pi/4,3\pi/4]$.
To make calculations tractable, we use the following approximate expressions of $u$ and $\psi$,
\begin{align}
        \label{u_reduced}
        u &\approx 2\sqrt{2\left(1-\cos(\theta-\theta')\right)}, \\
        \label{psi_reduced}
        \cos2\psi & \approx \frac{\cos2\theta+\cos2\theta'-2\cos(\theta+\theta')}{2\left(1-\cos(\theta-\theta')\right)} = -\cos(\theta+\theta'),
\end{align}
that are obtained from Eqs.~\eqref{exp_u} and \eqref{exp_psi} under the approximation $r\approx r' \approx 2$, which is justified by the fact that the pair correlation function is non-zero essentially close to contact for small deformations.
After some algebra, one finds that the tensor $\gUpsilon_\theta'$ has a leading contribution at order $q^2$, and thus vanishes at order $q$.
Turning to the evaluation of the tensor $\gUpsilon_r'$, we get that its leading order contribution is of order $q$, and is given by
\begin{equation}
\gUpsilon_r' = - \frac{A}{9}\, B \rho^3 q \, \Msymt{1}{0},
\end{equation}
where the coefficient $B$ is given in Eq.~\eqref{eq:coef:B}.
Finally we use Eq.~\eqref{calcul_sigmaq} to eliminate $q$ in favor of $\gSigma'$, leading to Eq.~\eqref{def_Upsilon_exp}.

\end{document}